\documentclass[twocolumn]{aastex631}

\DeclareRobustCommand{\textsc}[1]{{\mdseries\scshape #1}}
\providecommand{\uat}[2]{\href{http://astrothesaurus.org/uat/#2}{#1 (#2)}}
\makeatletter
\@ifundefined{@keys@name}{}{%
  \def\@keys@name{\textit{Unified Astronomy Thesaurus concepts:}\/~\mbox{}}}
\@ifundefined{frontmatter@keys@format}{}{%
  \g@addto@macro\frontmatter@keys@format{\hangindent=0pt\hangafter=0\relax}}
\makeatother

\makeatletter
\def\sec@upcase#1{#1}
\def\abstractname{Abstract}
\makeatother
\usepackage{etoolbox}
\patchcmd{\thebibliography}{REFERENCES}{References}{}{\PackageWarning{main}{References heading patch failed}}
\makeatletter
\def\frontmatter@RRAP@format{\small\centering\everypar{\hbox\bgroup}\def\par{\@ifvmode{}{\unskip\egroup\@@par}}}
\makeatother

\usepackage{verbatim}
\usepackage{amsmath,amstext,mathrsfs}
\usepackage{afterpage}    
\usepackage{marginnote}
\usepackage{makecell}
\usepackage{xfrac}
\usepackage{bm}
\makeatletter\let\old@foot@note@text\footnotetext \let\old@foot@note@mark\footnotemark\makeatother

\defcitealias{LazarianPogosyan2016}{LP16}
\defcitealias{LazarianPogosyan2012}{LP12}
\newcommand{\LP}{\citetalias{LazarianPogosyan2016}}

\shorttitle{Recovering 3D Magnetic Turbulence from Single-frequency Faraday Screens}
\shortauthors{Melnichenka, Lazarian, \& Pogosyan}
\received[Received ]{2026 June 1}
\revised[revised ]{2026 September 19}
\accepted[accepted ]{2026 September 22}
\published[published ]{2026 October 21}

\journalinfo{Published in The Astrophysical Journal, 1011:16 (2026), doi:10.3847/1538-4357/aeac22}

\begin{document}

\title{Recovering 3D Magnetic Turbulence from Single-frequency Faraday Screens}

\author[0009-0004-1969-779X]{A. Melnichenka}
\affiliation{Department of Physics, Berea College, Berea, KY 40404, USA}

\author[0000-0002-7336-6674]{A. Lazarian}
\affiliation{Department of Astronomy, University of Wisconsin--Madison, Madison, WI 53706, USA}

\author[0000-0002-7998-6823]{D. Pogosyan}
\affiliation{Department of Physics, University of Alberta, Edmonton, AB T6G 2E1, Canada}

\begin{abstract}

Statistics of polarized synchrotron radiation carry information about the properties of the underlying turbulence.  We consider a setup in which synchrotron radiation is emitted in a distant volume and then passes through a turbulent screen that induces Faraday rotation. Using both MHD simulations and synthetic turbulence spectra, we explore the spectrum of observed polarization directions measured at a single frequency as a diagnostic for recovering the statistics of turbulence in both the emitting region and the Faraday rotation volume. We compare these results with our analytical expectations. In doing so, we compare the directional spectrum with the wavelength-derivative diagnostic explored earlier by Lazarian \& Pogosyan. We demonstrate that the directional spectrum exhibits greater sensitivity to the turbulence in the Faraday rotation volume. We provide an observer-friendly criterion to determine whether it samples turbulence in the synchrotron-emitting region or in the Faraday rotation region. These results open a practical pathway for extracting turbulence statistics from existing and forthcoming single-band radio polarimetry.

\end{abstract}


\section{Introduction}
\label{sec:intro}

Magnetic turbulence is ubiquitous in astrophysics \citep{Elmegreen2004, Brandenburg2013}. In the interstellar medium alone, a Kolmogorov-like spectrum of density fluctuations extends over many decades in scale \citep{Armstrong1995}. Magnetic turbulence affects star formation \citep{McKee2007} and angular momentum transport \citep{Balbus1991, Balbus1998} and mediates heat conduction and cosmic-ray propagation \citep{Brandenburg2013}, yet most observational diagnostics are limited to 2D projections along the line of sight \citep[LOS;][]{Elmegreen2004}. Faraday rotation is one of the most widely used probes of these magnetized plasmas. From Galactic rotation measure (RM) grids and pulsars to cluster radio sources and fast radio bursts, it reveals the LOS magnetic field weighted by the thermal electron density, and through its spatial statistics, encodes the underlying turbulence \citep[e.g., see the review by][]{Akahori2018}.

The Faraday dispersion function formalism of \citet{Burn1966} and the RM synthesis technique of \citet{Brentjens2005} have greatly increased the amount of information that can be extracted from polarization observations, as they allow polarized emission to be separated by the degree of Faraday rotation it has experienced.

Recent work has tested synchrotron-based turbulence diagnostics numerically, validating intensity and polarization statistics on synthetic and MHD simulation data \citep[e.g.,][]{Herron2016, Lee2016, Zhang2016, Zhang2018, Zhang2023, Guo2024, Xiao2025, Zhang2025}, while single-frequency synchrotron intensity maps have also been used to trace the magnetic field direction through intensity gradients \citep{Lazarian2017}. 
In particular, \citet{Zhang2016} tested the multifrequency polarization-variance measures of \citet[][hereafter \LP]{LazarianPogosyan2016} and found that, when the random field dominates the Faraday rotation, the variance follows a universal $\lambda^{-2}$ law that carries no information on the inertial-range slope, whereas a strong mean LOS field restores the sensitivity of these measures to the slope.

Earlier work used synchrotron intensity and polarized intensity fluctuations to infer the spectrum and anisotropy of magnetic turbulence. Observational data were available, but the theoretical framework for interpreting them was provided by \citet[][hereafter LP12]{LazarianPogosyan2012}, who derived synchrotron statistics for an arbitrary electron index and, crucially, adopted a physically motivated model of anisotropic MHD turbulence consistent with \citet{GoldreichSridhar1995}.

Building on this, \LP{} incorporated Faraday rotation and established how synchrotron polarization statistics depend on the underlying magnetic and electron-density fluctuations along the LOS.
\LP{} developed position--position--frequency statistics of polarized synchrotron emission and introduced Faraday-sensitive measures that involve spectral derivatives (e.g., correlations of $dP/d\lambda^{2}$). Such diagnostics are powerful but require accurate multifrequency sampling to form derivatives, and they operate on the complex polarization $P(\bm X,\lambda^2)$ rather than directly on the polarization angle.

Here, we investigate a new use of the structure function for the polarization angle $\chi$, in the $\pi$-periodic form written through normalized Stokes parameters by \citet{Lazarian2022}, namely,
\begin{equation}
D_\chi = \frac{1}{2} \Big\langle 1 - \cos 2\big[ \chi(\bm{X},\lambda^2)- \chi(\bm{X}+\bm{R},\lambda^2)\big]\Big\rangle,
\label{eq:Dchi}
\end{equation}
where the 2D vector $\bm{R}$ is the separation on the sky between the two lines of sight;
throughout the rest of the paper we work with the rescaled form $D_u = 2\,[1-\langle\cos 2\Delta\chi\rangle] = 4\,D_\chi$.\footnote{$D_\chi$ differs from the polarization-angle dispersion function introduced by \citet{Hildebrand2009} and used in the form $1-\langle\cos\Delta\Phi\rangle$ by \citet{Houde2009} primarily in the argument of the cosine, which contains an additional factor of 2 to reflect the $\pi$-periodicity of the polarization angle $\chi$. We also include a normalization factor of $1/2$.
}

\begin{figure}[!t]
  \centering
  \includegraphics[width=\columnwidth]{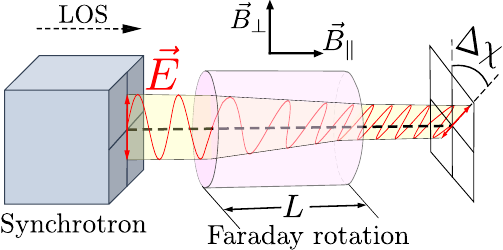}
  \caption{Schematic of the separated-screen geometry adopted in this work. Polarized synchrotron emission (gray cube), produced in a background region (e.g., a galactic halo with low thermal-electron density), passes through a foreground Faraday-rotating screen (pink cylinder), where most of the RM accumulates. Synchrotron emission and Faraday rotation are spatially separated. As shown, the polarization rotation due to the Faraday effect is represented by a change in the polarization angle, $\Delta \chi$, which depends on wavelength.}
  \label{fig:schematics}
\end{figure}

Angle dispersion and angle structure function statistics have a long history in polarimetry as turbulence diagnostics and as the basis of Davis--Chandrasekhar--Fermi-type \citep{Davis1951, Chandrasekhar1953} plane-of-sky field estimates, including modern structure function and ``dispersion function'' refinements, some of which treat beam and LOS averaging \citep[e.g.,][]{FalcetaGoncalves2008, Hildebrand2009, Houde2009, Houde2011}.
In this paper, we use this statistic to study the possibility of recovering the scaling properties of MHD turbulence from a synchrotron signal that is subjected to foreground Faraday rotation. Our setup is schematically illustrated in Figure~\ref{fig:schematics}.

We derive a Fourier-spectrum estimator, complementary to the angle structure function, from $\cos 2\chi$ and $\sin 2\chi$ maps obtained from normalized Stokes parameters.  We call the complex map, $u(\bm{X}) = \cos 2\chi(\bm{X}) + i \sin 2\chi(\bm{X})$, the ``director'' field.
We then investigate the parameter space where this estimator is most robust and where limited additional frequency coverage is required. 

In the Faraday screen setting, we show numerically that our director-field statistic $D_u(R;\lambda)$ recovers the inertial-range properties of the underlying 3D magnetic turbulence even at a single frequency, without requiring the wavelength array needed for the derivative measures of \LP{} (see the single-frequency angle statistics of \citealt{Lazarian2022}, obtained in the absence of Faraday rotation, and the single-frequency polarization-gradient spectra of \citealt{Zhang2025}).

In this paper, we do not address the more complex case of internal Faraday depolarization in the emitting region and coupled synchrotron emissivity and Faraday fluctuations, which is harder to interpret and invert (\citealt{Burn1966}; \citealt{Tribble1991}; \citealt{Sokoloff1998}, and \citetalias{LazarianPogosyan2016}). The separated-screen geometry on which we focus, the limiting case treated in Appendix~C of \LP{}, is still applicable to many astrophysical settings,  e.g., extragalactic background sources seen through the Milky Way \citep{Taylor2009, Jansson2012, Oppermann2012, Beck2015, Haverkorn2015}.

In Section~\ref{sec:theory}, we introduce the polarization angle director field and define the corresponding angle-based correlation and structure functions, emphasizing their sensitivity to Faraday rotation. We also show how the same statistic can be computed efficiently in Fourier space as a directional power spectrum, and summarize the expected inertial-range scaling and two-regime behavior. In Section~\ref{sec:numerics}, we describe the numerical setup based on both \textsc{AthenaK} MHD simulations and synthetic Faraday screens. We present results from the \textsc{AthenaK} runs in Section~\ref{sec:results_athena} and from the synthetic screens in Section~\ref{sec:results_synthetic}. In Section~\ref{sec:comparison_separated_screen}, we compare our measure with other polarization diagnostics, including \LP{} polarization and derivative-based polarization measures, and discuss the regimes where each method most robustly recovers the inertial-range slope. We discuss and summarize our results in Section~\ref{sec:conclusions}.

\section{Statistics of Polarization}
\label{sec:theory}

\subsection{Observables and the Director Field}
\label{subsec:observables}

We work with Stokes maps $Q(\bm X,\lambda^2)$ and $U(\bm X,\lambda^2)$ at one observing wavelength~$\lambda$.
The polarization angle is 
\begin{equation}
    \chi(\bm X,\lambda^2)=\tfrac12\arg(Q+iU) ~,
\end{equation}
defined modulo~$\pi$.
To respect the $\pi$-periodicity we introduce the unit director field
\begin{equation}
u(\bm X,\lambda^2)= e^{2i\chi(\bm X,\lambda^2)}
=\frac{Q+iU}{\sqrt{Q^2+U^2}}.
\label{eq:director}
\end{equation}
Only circular differences enter through $u(\bm X)\,u^\ast(\bm X+\bm R)=e^{2i\Delta\chi}$, so no angle unwrapping is required.

We consider a polarized background with intrinsic complex polarization~$P_i=|P_i| e^{2 i \psi}$ passing through a foreground magnetized slab of thickness~$L$.
The Faraday RM in CGS units is given by \citep{Burn1966}
\begin{equation}
\Phi(\bm X)=\frac{e^3}{2\pi m_e^2 c^4}\int_{0}^{L}n_e(\bm X,z)\,B_\parallel(\bm X,z)\,dz\,,
\label{eq:Phi_def}
    \end{equation}
and for a source without internal Faraday rotation or beam depolarization \citep{Brentjens2005}, the angle of the observed polarization is 
\begin{equation}
\chi(\bm X,\lambda^2)=\psi(\bm{X})+\lambda^{2}\Phi(\bm X)\,,
\label{eq:chi_obs}
\end{equation}
where $\psi(\bm{X})$ is the phase of the source signal. Throughout this paper we adopt the flat-sky (parallel-lines-of-sight) approximation, appropriate for a compact region observed at large distance, so that the 2D position $\bm X$ and separation $\bm R$ are transverse coordinates with length dimensions rather than angles. They map to an angular separation only through $\theta=R/d$ for a source at distance $d$, which we leave unspecified. All statistics below depend on separation only through the dimensionless ratios $R/r_i$ and $R/r_\phi$ (and, in Fourier space, through $kL$), so the results are independent of whether $\bm R$ is expressed in length or angular units.

Table~\ref{tab:symbols} lists the symbols used throughout the paper.
\begin{table*}[t]
\tabletypesize{\scriptsize}
\centering
\caption{Parameters for the Correlation Studies of Synchrotron Polarization in Separated Emitting and Faraday-rotating Regions}
\label{tab:symbols}
\setlength{\tabcolsep}{4pt}
\renewcommand{\arraystretch}{1.1}
\begin{tabular}{lll}
\hline\hline
Parameter & Meaning & First Appearance \\
\hline
$\bm X,\bm R$ & 2D sky position; 2D separation in the flat-sky approximation & Equation~(\ref{eq:Dchi}) \\
$\chi(\bm X,\lambda^2)$;\; $u=e^{2i\chi}$ & Polarization angle (mod $\pi$); unit director field & Equations~(\ref{eq:Dchi}), (\ref{eq:director}) \\
$P_i(\bm X)$;\; $\Phi(\bm X)$;\; $\phi(\bm X,z)$ & Intrinsic complex polarization; Faraday RM; RM density & Equations~(\ref{eq:chi_obs}), (\ref{eq:Phi_def}); Section~\ref{subsec:faraday} \\
$\xi_u(R;\lambda)$;\; $D_u(R;\lambda)$ & Director correlation and structure function & Equations~(\ref{eq:xiu_def}), (\ref{eq:Deta_def}) \\
$P_u(\bm k,\lambda)$ & Directional power spectrum & Equation~(\ref{eq:Pu_def}) \\
$\mathcal{M}(k)$ & Smoothed spectral proxy & Equation~(\ref{eq:Mk_def}) \\
$D_\psi(R)$;\; $D_\Phi(R)$ & Intrinsic phase; RM structure functions & Equation~(\ref{eq:xiu_factorized}) \\[2pt]
\multicolumn{3}{@{}l@{}}{Scales:}\\
$\lambda$;\; $L$ & Observing wavelength; LOS depth of Faraday screen & Equations~(\ref{eq:Dchi}), (\ref{eq:Phi_def}) \\
$r_i$;\; $r_\phi$ & Correlation lengths of intrinsic polarization and RM density & Equations~(\ref{eq:Dpsi_reg}), (\ref{eq:DPhi_inertial}) \\
$R_\times$;\; $k_\times$ & Domain-boundary separation and wavenumber & Equations~(\ref{eq:Rtimes}), (\ref{eq:ktimes}) \\[2pt]
\multicolumn{3}{@{}l@{}}{Spectral indices:}\\
$m_i$;\; $m_\phi$;\; $m_\Phi$ & Intrinsic (2D); RM density (3D); RM (2D, $m_\Phi=1+m_\phi$) exponents & Equations~(\ref{eq:Dpsi_reg}), (\ref{eq:DPhi_inertial}), (\ref{eq:DPhi_reg}) \\[2pt]
\multicolumn{3}{@{}l@{}}{Basic statistics:}\\
$\eta=2\sigma_\Phi \lambda^2$ & Faraday rotation strength (twice the rms rotation angle) & Equations~(\ref{eq:du}), (\ref{eq:faraday_rotation_strength}) \\
$\sigma_\psi$;\; $\sigma_\phi$;\; $\sigma_\Phi$ & rms intrinsic polarization phase; rms RM density; rms RM (saturated) & Equations~(\ref{eq:Dpsi_reg}), (\ref{eq:DPhi_inertial}), (\ref{eq:DPhi_reg}) \\
\hline
\end{tabular}
\end{table*}

\subsection{Directional Correlation and Angle Structure Function}
\label{subsec:strfn}

Our basic two-point statistic is the directional correlation,
\begin{equation}
\xi_u(\bm R;\lambda)
=\langle u(\bm X,\lambda^2)\,u^\ast(\bm X+\bm R,\lambda^2)\rangle
=\langle \cos 2\Delta\chi\rangle,
\label{eq:xiu_def}
\end{equation}
and its  structure function form
\begin{equation}
D_u(\bm{R};\lambda)= 2 [1-\xi_u(R;\lambda)]\,.
\label{eq:Deta_def}
\end{equation}

From Equation~(\ref{eq:chi_obs}), $u(\bm X,\lambda^2)=e^{2i\psi(\bm X)}\,e^{2i\lambda^2\Phi(\bm X)}$.
Assuming the intrinsic polarization direction and the Faraday screen are statistically independent, the director correlation factorizes
\begin{equation}
\xi_u(\bm{R};\lambda)
=\exp\bigl[-2 D_\psi(\bm{R})\bigr]\;\exp\bigl[-2\lambda^{4}\,D_\Phi(\bm{R})\bigr]\,,
\label{eq:xiu_factorized}
\end{equation}
where 
\begin{equation}
D_\psi(\bm{R})=\left\langle\left[\psi(\bm{X}+\bm{R})-\psi(\bm{X})\right]^2\right\rangle 
\end{equation}
is the structure function of the intrinsic polarization phases at the source and
\begin{equation}
D_\Phi(\bm{R})=\left\langle\left[\Phi(\bm{X}+\bm{R})-\Phi(\bm{X})\right]^2\right\rangle
\end{equation}
is the RM structure function.
Here we have assumed Gaussian statistics for $\chi$ and $\Phi$. 

In the special case of a uniform polarized background, this reduces to
\begin{equation}
D_u(\bm{R};\lambda)
=2\Bigl[1-e^{-2\lambda^{4}\,D_\Phi(\bm{R})}\Bigr]\,.
\label{eq:Deta_gauss}
\end{equation}
In the weak-rotation limit $2\lambda^{4}D_\Phi\ll1$, this linearizes to
$D_u(\bm{R};\lambda)=4 \lambda^{4}\,D_\Phi(\bm{R})$,
so the angle structure function directly traces the RM structure function, scaled by $\lambda^4$.

\subsection{Intrinsic Polarization Model}
\label{subsec:intrinsic}

We write the intrinsic polarization as $P_i(\bm X)=\bar P_i\,e^{i\bar\psi}+\delta P_i(\bm X)$ with mean $\bar P_i$ and zero-mean fluctuation $\delta P_i$. The 2D projected correlation of $\delta P_i$ has correlation length~$r_i$ and inertial-range index~$m_i$.

Following \LP{}, we adopt a regularized isotropic model for the intrinsic polarization-angle correlations.
\citet{Lazarian2022} showed that if the mean polarization dominates the fluctuations, as in a sub-Alfv\'enic regime, then the statistics of the polarization angle follow the scaling of the polarization itself. 
Thus, if we adopt the model
\begin{equation}
D_\psi(R)
=2\,\sigma_\psi^2\,\frac{(R/r_i)^{m_i}}{1+(R/r_i)^{m_i}}
\label{eq:Dpsi_reg}
\end{equation}
the spectral index $m_i$ will be similar to that of the magnetic field, but projected onto the 2D plane.
If the 2D source polarization is accumulated through a turbulent region of significant depth, the Kolmogorov scaling corresponds to $m_i=1+2/3=5/3$, where the extra unity comes from integrating the 3D field along the LOS (\citetalias{LazarianPogosyan2012}). Here we treat $m_i$ as a free parameter, thus extending the applicability of the ansatz Equation~(\ref{eq:Dpsi_reg}). The intrinsic phase rms value $\sigma_\psi$ has the upper bound $\sigma_\psi < \pi/\sqrt{12}$, which is approached in the limit of highly super-Alfv\'enic turbulence, where the polarization directions become nearly uniformly distributed.
It is proportional to the Alfv\'enic Mach number, $\sigma_\psi \propto M_A \ll 1 $, in the sub-Alfv\'enic regime where a strong regular magnetic field sets the preferred direction.

\subsection{RM Model}
\label{subsec:faraday}

Let $\phi(\bm X,z) \propto n_e(\bm X,z)\,B_\parallel(\bm X,z)$ be the 3D RM density field, with rms~$\sigma_\phi$, correlation length~$r_\phi$, and correlation index~$m_\phi$, such that $\Phi(\bm X)=\int_0^L\phi(\bm X,z)\,dz$. The correlation index $m_\phi$ is related to the 3D spectral index as $P_\phi(k) \propto k^{n_\phi} \propto k^{-3-m_\phi}$, so positive $m_\phi>0$ corresponds to steep spectra, while negative $m_\phi<0$ corresponds to shallow RM-density spectra.  The RM-density spectrum is expected to be steep when it is dominated by the fluctuations of $B_\parallel$ on a background of relatively uniform density of thermal electrons (Kolmogorov scaling is $m_\phi = 2/3$, as observed for the electron density of the diffuse ionized medium by \citealt{Armstrong1995}), but can become shallow when density fluctuations are important, as happens in supersonic turbulence where the density variance is concentrated in shock-compressed structures (\citetalias{LazarianPogosyan2016}; \citealt{Xu2021}; \citealt{Xiao2025}). 

For a {thick} Faraday screen ($L\gg r_\phi$), \LP{}  show that for $R\ll r_\phi$ (their Equation (148), with $0<m_\phi<1$),
\begin{equation}
D_\Phi(R)
=2\sigma_\phi^2\,L\,r_\phi^{-m_\phi}\;R^{\,1+m_\phi}\,,
\label{eq:DPhi_inertial}
\end{equation}
so the RM structure function scales as $D_\Phi(R)\propto R^{1+m_\phi}$ in this regime. The order-unity numerical constant, which depends on the shape of the RM-density correlation function, is absorbed into the definition of $r_\phi$.
We will use the regularized ansatz
\begin{equation}
D_\Phi(R)
=2\sigma_\Phi^2\,\frac{(R/r_\phi)^{m_\Phi}}{1+(R/r_\phi)^{m_\Phi}}\,,
\label{eq:DPhi_reg}
\end{equation}
where $\sigma_\Phi^2$ is the variance and $m_\Phi$ is the scaling index for the 2D RM field. Comparing Equation~(\ref{eq:DPhi_reg}) with Equation~(\ref{eq:DPhi_inertial}) at $R\ll r_\phi$ gives $\sigma_\Phi^2=\sigma_\phi^2\,L\,r_\phi$ and $m_\Phi=1+m_\phi$, with integration along the LOS raising the correlation index by one, exactly as it does for the intrinsic index $m_i=1+2/3$ of Section~\ref{subsec:intrinsic}. This ansatz can be used
for $0 < m_\Phi < 2$, which corresponds to the RM-density spectrum in the range $ -4 < n_\phi < -2 $. 

It is worth emphasizing which of these quantities are directly observable. The saturated RM dispersion $\sigma_\Phi$ is the standard deviation of the Faraday RM across the sky, and is measured directly from an RM map, for example by RM synthesis \citep{Brentjens2005} or by fitting the polarization angle against $\lambda^2$ over the available bands \citep[see, e.g., the RM maps of][]{Taylor2009, Oppermann2012}. The decomposition $\sigma_\Phi^2=\sigma_\phi^2\,L\,r_\phi$ reexpresses this observable in terms of three-dimensional quantities, the intrinsic RM-density variance $\sigma_\phi^2$, the screen depth $L$, and the RM-density correlation length $r_\phi$, none of which is separately observable. The relation is useful for physical interpretation but is not needed to evaluate the diagnostics below, which depend on $\sigma_\Phi$ directly. The Faraday rotation strength $\eta=2\sigma_\Phi\lambda^2$ (Equation~(\ref{eq:faraday_rotation_strength})) is likewise an observable. It is twice the rms Faraday rotation angle, in radians, imprinted by the screen at wavelength $\lambda$, and is obtained by multiplying the measured RM dispersion by $2\lambda^2$. It sets the operating regime, $\eta\ll1$ weak (the single-band slope is directly readable), $\eta\sim1$ transitional, $\eta\gg1$ strong, so that an observer can compute $\eta$ from a measured RM dispersion and an observing wavelength alone, without recourse to $\sigma_\phi$, $L$, or $r_\phi$ individually. We refer to $\eta<1$ as moderate rotation.

\subsection{Weak-rotation Expansion and the Two-slope Regime}
\label{subsec:two_slope}

We now expand the factorized director correlation (Equation~(\ref{eq:xiu_factorized})) in the regime where both intrinsic decorrelation and Faraday rotation are small.
For weak decorrelation (small $D_\psi$ and small~$2\lambda^4 D_\Phi$), we expand each factor to leading order:

\begin{equation}
    e^{-2\lambda^4 D_\Phi} = 1-2\lambda^4 D_\Phi .
\end{equation}
Therefore,
\begin{equation}
    \xi_u=1-2 D_\psi -2\lambda^4 D_\Phi
\end{equation}
and hence
\begin{equation}
    D_u=4D_\psi+4\lambda^4 D_\Phi ~.
\end{equation}
Substituting \eqref{eq:Dpsi_reg} and \eqref{eq:DPhi_reg} we obtain the expanded form of the directional structure function 
\begin{equation}
    \frac{D_u}{2}=4\sigma_\psi^2 \,\frac{(R/r_i)^{m_i}}{1+(R/r_i)^{m_i}}+\eta^2\,\frac{(R/r_\phi)^{m_\Phi}}{1+(R/r_\phi)^{m_\Phi}}\,,
    \label{eq:du}
\end{equation}
where the dimensionless coefficient 
\begin{equation}
    \eta=2\sigma_\Phi \lambda^2
    \label{eq:faraday_rotation_strength}
\end{equation}
describes the Faraday rotation strength in terms of twice the rms rotation angle (in radians) through the Faraday region. 

In Equation~(\ref{eq:du}) the first term is the intrinsic (synchrotron-angle) contribution and the second is the Faraday contribution.
If $m_i \ne m_\Phi$ one can observe the characteristic scale $R_\times$ where the director structure function changes the slope.
If the transition happens in the inertial range, $R\ll r_i$ and $R\ll r_\phi$, Equation~(\ref{eq:du}) reduces to
\begin{equation}
\frac{D_u}{2}
\approx 4\sigma_\psi^2 \left(\frac{R}{r_i}\right)^{m_i}
+\eta^2\,\left(\frac{R}{r_\phi}\right)^{m_\Phi}
\label{eq:Du_inertial}
\end{equation}
and we can define $R_\times(\lambda)$ by the equality of the two terms in Equation~(\ref{eq:Du_inertial})
\begin{equation}
R_\times(\lambda)
=\left(\frac{4\sigma^2_\psi r_\phi^{\,m_\Phi}}{\eta^2\;r_i^{\,m_i}}\right)^{1/(m_\Phi-m_i)}.
\label{eq:Rtimes}
\end{equation}
Hence, $R_\times\propto\lambda^{-4/(m_\Phi-m_i)}$: increasing the observing wavelength shrinks the intrinsic-dominated domain and expands the Faraday-dominated one. Throughout this paper we focus on the case $m_i > m_\Phi$, which is the regime of primary astrophysical interest. The intrinsic synchrotron polarization typically follows a Kolmogorov-like cascade ($m_i \approx 5/3$ after LOS projection of a 3D Kolmogorov field), while the Faraday screen spectrum becomes shallower ($m_\Phi < m_i$) whenever density fluctuations dominate the RM, as occurs for supersonic turbulence (see the $M_s=10$ run in Section~\ref{sec:results_athena}). In this regime $m_\Phi - m_i < 0$, so $R_\times$ grows with wavelength and the Faraday-dominated branch progressively occupies larger and larger scales as $\lambda$ increases, providing a clean observational handle on the screen statistics.

The corresponding domain-boundary wavenumber follows from $k_\times\sim 1/R_\times$:
\begin{equation}
k_\times(\lambda)\propto\lambda^{\;4/(m_\Phi-m_i)}\,.
\label{eq:ktimes}
\end{equation}

\begin{figure*}[!t]
  \centering
  \begin{minipage}[t]{0.49\textwidth}
    \centering
    \includegraphics[width=\linewidth]{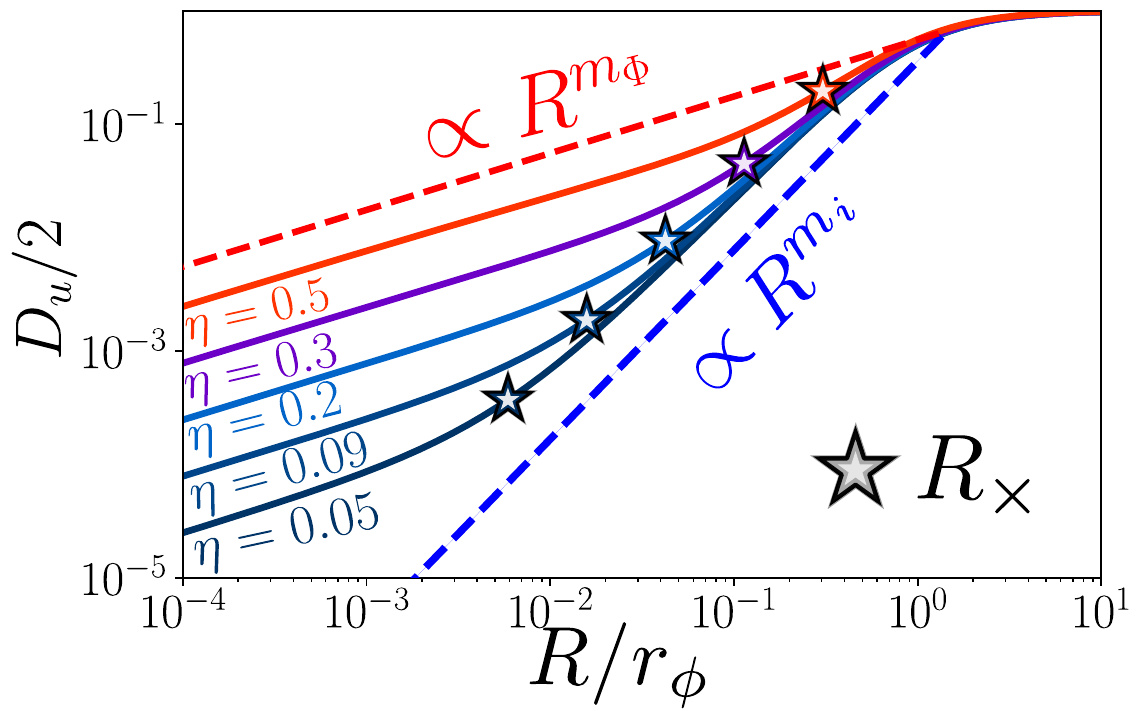}
  \end{minipage}\hfill
  \begin{minipage}[t]{0.49\textwidth}
    \centering
    \includegraphics[width=\linewidth]{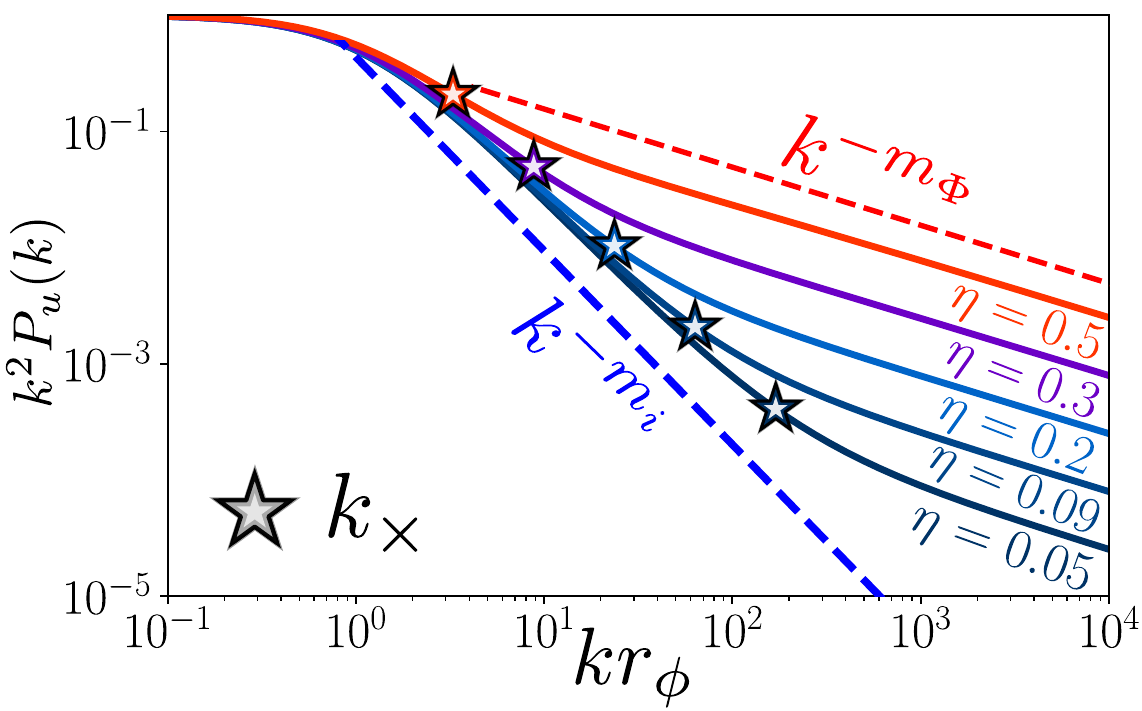}
  \end{minipage}\hfill
   \caption{The left panel shows the analytical expectation of the structure function $D_u(R)/2$, predicted by Equation~\eqref{eq:du}, as a function of separation $R$ for the density-only synthetic Faraday screen at several values of~$\eta$ (colored curves from $\eta=0.05$ in blue to $\eta=0.5$ in red). The example shows the astrophysically relevant case $m_i > m_\Phi$, with $m_i = 5/3$ (Kolmogorov synchrotron source) and $m_\Phi < m_i$ (density-dominated Faraday screen). Dashed lines indicate the intrinsic ($\propto R^{m_i}$, blue) and Faraday ($\propto R^{m_\Phi}$, red) inertial-range asymptotics of the ansatz \eqref{eq:Du_inertial}. The $\bigstar$ marks the crossover scales $R_\times$ and $k_\times$, predicted by Equations~\eqref{eq:Rtimes} and \eqref{eq:ktimes}. As $\eta$ varies, $R_\times$ shifts accordingly and lies at the transition between the Faraday-dominated ($\propto R^{m_\Phi}$) and intrinsic ($\propto R^{m_i}$) regimes.
The right panel shows the smoothed power-spectrum proxy $\mathcal{M}(k) \propto k^2 P_u(k)$ (Equation~\ref{eq:Mk_def}) for the same parameter set. Dashed lines show the corresponding $k^{-m_i}$ and $k^{-m_\Phi}$ scalings. The $\bigstar$ on the right panel denotes the crossover scale $k_\times$.}
  \label{fig:sf_demo_realspace}
\end{figure*}

Figure~\ref{fig:sf_demo_realspace} shows the structure function $D_u(R;\lambda)/2$ for this configuration at several values of~$\eta$.
There are two domains. At separations $R<R_\times$, the shallower of the two power laws dominates, and at $R>R_\times$, the steeper one takes over.
The crossover scale $R_\times$ shifts with $\eta$ according to Equation~(\ref{eq:Rtimes}).

\subsection{Directional Spectrum}
\label{subsec:dirspec}

Define the director components
\begin{equation}
n_1=\cos 2\chi\,,\qquad n_2=\sin 2\chi\,.
\end{equation}
Then $\xi_u(R;\lambda)=\langle n_1 n_1'\rangle+\langle n_2 n_2'\rangle$, where primes denote evaluation at $\bm X+\bm R$.
By the Wiener--Khinchin theorem, $\xi_u$ is the inverse Fourier transform of the {directional power spectrum}
\begin{equation}
P_u(\bm k,\lambda)
=|\widehat{n_1}(\bm k)|^2+|\widehat{n_2}(\bm k)|^2
=\left|\mathcal{F}\left[\frac{Q+iU}{\sqrt{Q^2+U^2}}\right]\right|^2.
\label{eq:Pu_def}
\end{equation}
Its ring average is $P_u(k,\lambda)$.
This implementation requires only a single fast Fourier transform (FFT) of the complex director map and avoids pairwise averaging.

The directional power spectrum $P_u(\bm k,\lambda)$ is the two-dimensional power spectrum of the polarization orientation with the polarized intensity divided out. It is the Fourier conjugate of the directional correlation $\xi_u$ (Equation~(\ref{eq:xiu_def})). The two carry identical information, but the spectrum isolates individual scales. It is well suited to the Faraday screen problem. First, because $u$ is amplitude normalized, $P_u$ is insensitive to intrinsic brightness fluctuations and to overall flux calibration, responding only to the spatial variation of the polarization angle. Second, because $u$ inherits the $\pi$-periodicity of the polarization angle (it is built from $2\chi$), no angle unwrapping is required. Third, a spatially uniform rotation, such as that produced by the mean LOS field of the screen, shifts every angle by the same amount and therefore contributes only to the $\bm k=0$ mode, leaving $P_u(k>0)$ unchanged. Consequently $P_u$ probes the fluctuating part of the Faraday screen, and its inertial-range slope reports the RM-fluctuation exponent $m_\Phi$ (through the $\eta^2 k^{-(m_\Phi+2)}$ term of Equation~(\ref{eq:Pu_asymptotics})), while the intrinsic synchrotron fluctuations set the complementary $m_i$ branch. Since $u$ is dimensionless, $P_u$ is defined up to the Fourier normalization and is quoted in arbitrary units throughout.

\subsection{Hankel Transform and Slope Identification}
\label{subsec:hankel}

To translate the real-space two-slope structure function (Equation~(\ref{eq:Du_inertial})) into Fourier space, we use the standard zero-order Hankel transform: $\int R^m J_0(kR)\,R\,dR\propto k^{-(m+2)}$, understood as a regularized transform valid for $0<m<2$.
The two power-law terms in Equation~(\ref{eq:Du_inertial}) therefore contribute
\begin{equation}
\hspace*{-6pt}\scalebox{0.91}{$\displaystyle\medmuskip=3mu\thickmuskip=4mu
P_u(k;\lambda)\propto
4\sigma_\psi^2 r_i^{-m_i}\,k^{-(m_i+2)}
+\eta^2\,r_\phi^{-m_\Phi}\,k^{-(m_\Phi+2)}
+\cdots$}
\label{eq:Pu_asymptotics}
\end{equation}
in the inertial $k$-range.\footnote{In the Faraday-dominated range the directional spectrum has the same slope $-(m_\Phi+2) = -(m_\phi+3) $ as the three-dimensional power spectrum of the RM density, because the LOS projection preserves the per-mode index (\citetalias{LazarianPogosyan2012}). The corresponding ring-summed spectrum $E_u=2\pi k\,P_u$ would go as $k^{-(2+m_\phi)}$.}
Therefore, if $m_\Phi<m_i$, the Faraday term ($\propto k^{-(m_\Phi+2)}$) is shallower and dominates at $k\gg k_\times$ (small scales), while the intrinsic term dominates at $k\ll k_\times$. If $m_\Phi>m_i$, the Faraday term dominates at $k\ll k_\times$ (large scales). The two terms of Equation~(\ref{eq:Pu_asymptotics}) are equal not at $k=1/R_\times$ but at $k_\times=C/R_\times$, because the transform of $R^m$ carries the coefficient $c(m)=2^{m+1}\pi m\,\Gamma(1+m/2)/\Gamma(1-m/2)$, so that $C=[c(m_\Phi)/c(m_i)]^{1/(m_\Phi-m_i)}$. This fixes the constant in Equation~(\ref{eq:ktimes}). A crossover exists only for $\eta<2\sigma_\psi$, because the two terms of Equation~(\ref{eq:du}) saturate at $4\sigma_\psi^2$ and $\eta^2$, and beyond that value the Faraday term dominates at all scales.

\subsection{Smoothed Power Spectrum Proxy $\mathcal{M}(k)$}
\label{subsec:proxy}

To identify inertial-range slopes in Fourier space for the synthetic-screen figures we use a spectrum proxy $\mathcal{M}(k)$ constructed directly from the real-space structure function that is plotted in Figure~\ref{fig:sf_demo_realspace}
\begin{equation}
\mathcal{M}(k)=
\int \frac{D_u(R;\lambda)}{2}\,
W_{\sigma_{\ln}}\big(\ln(kR)\big)\,d\ln R,
\label{eq:Mk_def}
\end{equation}
with a normalized Gaussian window in $\ln(kR)$,
\begin{equation}
W_{\sigma_{\ln}}(x)=\frac{1}{\sqrt{2\pi}\,\sigma_{\ln}}
\exp\left(-\frac{x^2}{2\sigma_{\ln}^2}\right),
\quad x=\ln(kR),
\label{eq:Wln_def}
\end{equation}
where $\sigma_{\ln}$ is the smoothing width.
Because $W_{\sigma_{\ln}}$ is localized around $kR= 1$, a real-space power law $\tfrac{D_u}{2}\propto R^{m}$ maps to
\begin{equation}
\mathcal{M}(k)\propto k^{-m}
\end{equation}
(up to a constant factor set by $\sigma_{\ln}$), so the two inertial-range branches in Equation~(\ref{eq:Du_inertial}) appear as the reference scalings
$k^{-m_i}$ and $k^{-m_\Phi}$ in Figure~\ref{fig:sf_demo_realspace}.
The shallower branch dominates at $k\gg k_\times$ and the steeper branch dominates at $k\ll k_\times$, with $k_\times\sim 1/R_\times$.
Because $\mathcal{M}(k)$ is defined over the logarithmic measure $d\ln R$, its index is shallower by~2 than that of the directional spectrum: $P_u\propto k^{-(m+2)}$ corresponds to $\mathcal{M}(k)\propto k^{-m}$ (Equation~(\ref{eq:Pu_asymptotics})).

\section{Numerical Setup}
\label{sec:numerics}

We test our analytical predictions with two complementary numerical approaches.
First, we run self-consistent MHD simulations with \textsc{AthenaK} to obtain physically realistic turbulent fields (Section~\ref{subsec:athena}).
Second, because the MHD simulations have a limited inertial range, we construct synthetic Faraday screens with prescribed spectral slopes on larger grids, giving us a wider dynamic range and full control over the intrinsic and Faraday exponents (Section~\ref{subsec:synthetic}).

\subsection{AthenaK MHD Simulations}
\label{subsec:athena}

The MHD datasets were generated with \textsc{AthenaK} \citep{Stone2026}, which solves the standard compressible ideal MHD equations under periodic boundary conditions and an isothermal equation of state. The sound speed is set to $c_s=1$ (\hyperref[tab:sim]{A0}) and $c_s=0.1$ (\hyperref[tab:sim]{A1}), so that identical driving yields $M_s=1$ and $M_s=10$.
Turbulence is driven in the band $kL_{\rm box}/2\pi = 2$--$3$ with correlation time $t_{\rm corr}=0.24$.
The computational domain is a $1024^3$ periodic cube with $L_{\rm box}=1$.
The forcing is the Ornstein--Uhlenbeck module of \textsc{AthenaK} \citep[a stochastic forcing with a finite correlation time,][]{Eswaran1988, Federrath2010}, as in the runs of \citet{Hu2024}. Unlike that two-fluid study, the gas is treated here as a single ideal MHD fluid. We analyze two runs spanning transonic ($M_s=1$, run \hyperref[tab:sim]{A0}) to highly supersonic ($M_s=10$, run \hyperref[tab:sim]{A1}) conditions at fixed sub-Alfv\'enic $M_A=0.8$, summarized in Table~\ref{tab:sim}. The measured, volume-weighted states are $M_s=1.09$, $M_A=0.87$ for run A0 and $M_s=10.5$, $M_A=0.84$ for run A1. Figure~\ref{fig:dirspec_athena} averages run A0 over four saturated snapshots and run A1 over three. Earlier A1 dumps were excluded because their magnetic spectrum was still relaxing.

\begin{table}
\makeatletter
\def\tableft@skip{\@flushglue}\def\tabright@skip{\@flushglue}%
\appdef\tabular@hook{\def\@halignto{to\hsize}}%
\makeatother
\centering
 \begin{tabular}{ | c | c | c | c | c |}
		\cline{1-5}
		Run & $M_s$ & $M_A$ & $\beta$ & Resolution \\ \cline{1-5}\noalign{\vskip\doublerulesep}\cline{1-5}
		A0 & 1.0 & 0.8 & 1.28 & $1024^3$ \\ 
		A1 & 10.0 & 0.8 & 0.0128 & $1024^3$ \\
        \cline{1-5}
	\end{tabular}
	\caption{\label{tab:sim} Setups of MHD Turbulence Simulations}
	\vspace{2pt}
	{\raggedright\footnotesize \textit{Note.} The sonic and Alfv\'en Mach numbers, i.e., $M_s$ and $M_A$, are the target values, the measured volume-weighted values are given in the text. $\beta = 2(M_A/M_s)^2$ is the plasma beta.\par}
\end{table}

The sub-Alfv\'enic condition ($M_A<1$) ensures a strong mean magnetic field, so that the intrinsic polarization phase is well described by the linearized regime $|\delta P_i|\ll|\bar P_i|$ discussed in Section~\ref{subsec:intrinsic}.

For the analysis we take the LOS to be \emph{perpendicular} to the mean field $B_0$ that, correspondingly, lies in the plane of the sky. This is the geometry assumed in Section~\ref{subsec:intrinsic}, where the intrinsic polarization has a dominant mean about which the director is linearized, and it is also the geometry of the synthetic cubes of Section~\ref{subsec:synthetic}. 
From each snapshot we construct the intrinsic complex polarization by integrating the transverse magnetic field along the LOS 
\begin{equation}
P_i(\bm X)\propto\sum_z (B_x+iB_y)^2\,\Delta z\,,
\label{eq:Pi_athena}
\end{equation}
where $B_x$ and $B_y$ are the field components transverse to LOS.\footnote{We refer to the demonstration in \LP{} that using
a cosmic-ray index $\gamma \ne 2$ only weakly affects the correlation scaling of the synchrotron polarization.}
The Faraday RM is constructed from the front portion of the box (the ``screen,'' occupying the first quarter of the LOS depth):
\begin{equation}
\Phi(\bm X)=\kappa\sum_{z=0}^{z_{\rm scr}} n_e(\bm X,z)\,B_\parallel(\bm X,z)\,\Delta z\,,
\label{eq:Phi_athena}
\end{equation}
where $\kappa$ is a normalization constant.
The emission region extends over the remaining three-quarters of the box.
The observed polarization is $P(\bm X,\lambda^2)=P_i\,e^{i\,\eta\,\hat\Phi}$, where $\hat\Phi=\Phi/\sigma_\Phi$ is the RM map normalized to unit variance and $\eta= 2\lambda^2\sigma_\Phi$ is the dimensionless rotation parameter (labeled ``$\eta$'' in the figures).
We then form the director field $u=P/|P|$ and compute both $D_u(R;\lambda)$ and $P_u(k;\lambda)$ via Equation~(\ref{eq:Pu_def}).

\subsection{Synthetic Faraday Screens}
\label{subsec:synthetic}

To overcome the limited inertial range of the MHD runs, we generate synthetic turbulent cubes on $5040^3$ grids with prescribed spectral slopes, allowing us to control $m_i$ and $m_\Phi$ independently and verify the analytical predictions over a wider dynamic range.

The density field is generated as a Gaussian random field in Fourier space with a prescribed 3D power spectrum, $P_n(k)\propto k^{-2.26}$ for the $M_s=10$ case and $P_n(k)\propto k^{-11/3}$ for the $M_s=1$ case (close to the density spectra measured in the respective \textsc{AthenaK} runs, Figure~\ref{fig:dirspec_athena}), with $\langle n_e\rangle=1$ and $\sigma_n$ matching the statistics of the \textsc{AthenaK} simulation. We use this field as generated rather than exponentiating it into a lognormal one, because the exponentiation adds a shallow small-scale tail to the density spectrum that is absent from the prescribed power law. For $M_s=10$ the field is then not positive everywhere, which is immaterial here since only the LOS integrals of $n_e$ and $n_e B_\parallel$ enter.  
The perpendicular magnetic field components ($B_x$, $B_y$) are generated as independent Gaussian random fields with Kolmogorov power spectrum $P_B(k)\propto k^{-11/3}$, and a strong mean field $\bar B_x$ is added to set $M_A<1$. The LOS is the $z$-axis, so $\bar B_x$ is perpendicular to it and lies in the plane of the sky, the same geometry as in the \textsc{AthenaK} analysis of Figure~\ref{fig:dirspec_athena}. The parallel field in the screen is built from a separate Gaussian field $\delta B$ with the same $k^{-11/3}$ spectrum and a uniform $B_0=2.3\,\sigma_B$, with $\sigma_B$ the rms of $\delta B$, taken alone or together as described for the three columns of Figure~\ref{fig:dirspec_model} in Section~\ref{sec:results_synthetic}.
We consider two synthetic cases whose fluctuation amplitudes equal those measured in the last snapshots of the two \textsc{AthenaK} runs, namely 
$\bar n/\sigma_n=0.51$ and $\bar B_\perp/\sigma_{B_\perp}=3.6$ for $M_s=10$ and $\bar n/\sigma_n=2.37$ and $\bar B_\perp/\sigma_{B_\perp}=2.6$ for $M_s=1$. In the \textsc{AthenaK} runs $\sigma_{B_\perp}$ is the rms of the sky-plane field component perpendicular to $B_0$, the one that sets the polarization angle, and the synthetic cubes use this $\sigma_{B_\perp}$ for both sky-plane components.
From these cubes we construct $P_i(\bm X)$ using the same kernel as Equation~(\ref{eq:Pi_athena}), and the Faraday RM $\Phi(\bm X)=\kappa\sum_z n_e B_\parallel\,\Delta z$.

\section{Results of AthenaK MHD Simulations}
\label{sec:results_athena}
\begin{figure*}[t]
  \centering

  \includegraphics[width=\textwidth]{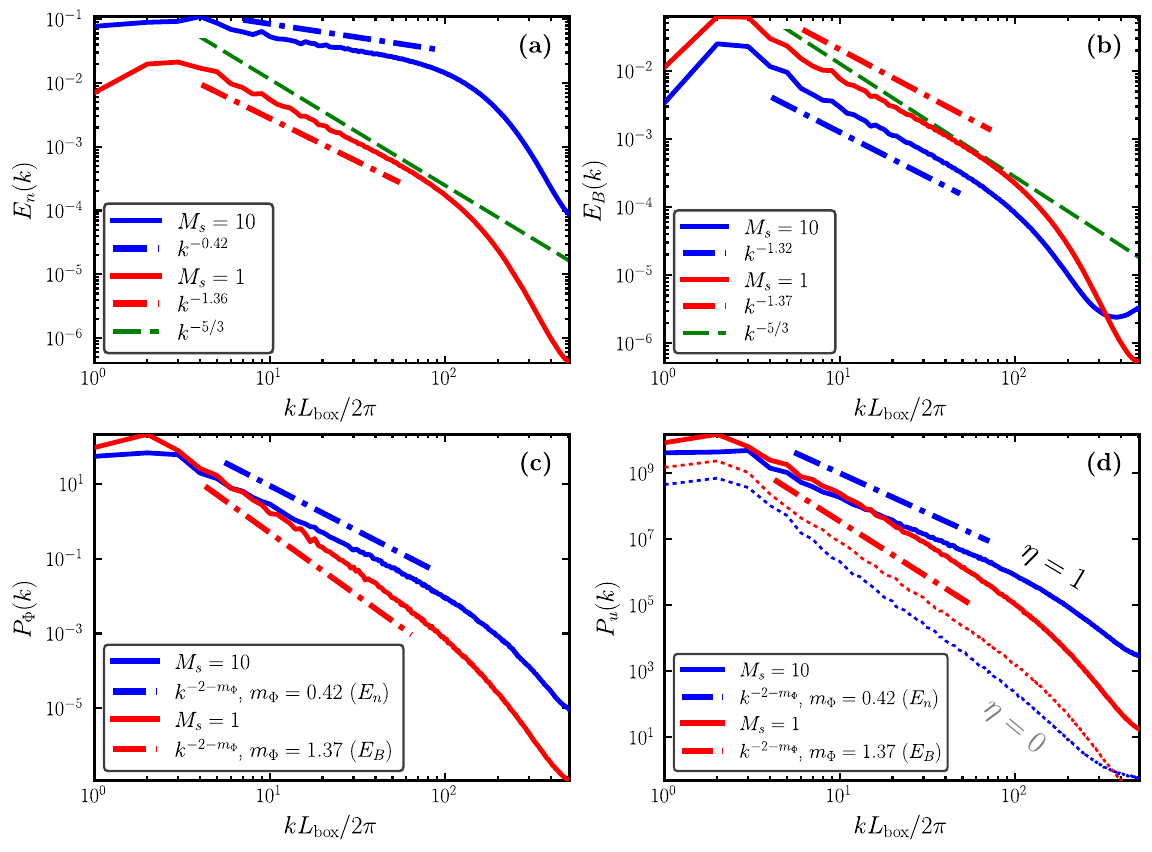}

  \caption{\textsc{AthenaK} $1024^3$ runs at $M_s=1$ (red) and $M_s=10$ (blue). All lines of a run share its color. $B_0\perp{\rm LOS}$, so that $B_0$ lies in the plane of the sky (Section~\ref{subsec:athena}). Curves are means over four ($M_s=1$) and three ($M_s=10$) saturated snapshots (Section~\ref{subsec:athena}). $E(k)$ is summed over a 3D shell or 2D ring and $P(k)$ is averaged over it. All dashed--dotted lines are offset from the curves for clarity. In (a) and (b) they are least-squares fits and extend beyond the fitted range. Panel (a) shows the 3D density spectrum $E_n$, with fits of $k^{-1.36}$ ($M_s=1$) and $k^{-0.42}$ ($M_s=10$). Panel (b) shows the 3D magnetic energy spectrum $E_B$ with the mean field excluded, with fits of $k^{-1.37}$ and $k^{-1.32}$. The dashed green line is the Kolmogorov $k^{-5/3}$ reference. Panel (c) shows the 2D spectrum $P_\Phi$ of the RM $\Phi=\int n_e B_\parallel\,dz$ through the front quarter of the box, which acts as the screen. Panel (d) shows the directional spectrum $P_u$ of Equation~\eqref{eq:Pu_def} at $\eta=1$ (solid) and $\eta=0$ (dotted). In (c) and (d) the dashed--dotted lines are not fits. Their slope is fixed at the prediction $k^{-(2+m_\Phi)}$ of Equation~\eqref{eq:Pu_asymptotics}, where $m_\Phi$ is read from the top row as minus the slope of the field that dominates the screen, $E_B$ for $M_s=1$ and $E_n$ for $M_s=10$. This gives $m_\Phi=1.37$ and $0.42$, and only the amplitude is fitted. Beyond $kL_{\rm box}/2\pi\approx 80$ all curves enter the numerical dissipation range.}
\label{fig:dirspec_athena}
\end{figure*}

Figure~\ref{fig:dirspec_athena} shows the directional spectrum $P_u(k;\lambda)$ from the two \textsc{AthenaK} runs of Table~\ref{tab:sim}
with significant Faraday rotation $\eta=1$.  The results are shown alongside the three-dimensional density and magnetic energy spectra, the latter summed over all three components of the fluctuating field, $E_B(k)=\tfrac12\sum_{i=x,y,z}|\delta B_i|^2(k)$, with the uniform mean field excluded, and the spectrum of the RM, which is two dimensional.\footnote{$E=4\pi k^2P$ in 3D and $E=2\pi kP$ in 2D, so Kolmogorov scaling reads $k^{-5/3}$ for the shell-summed $E$ of panels (a), (b) and $k^{-11/3}$ for the ring-averaged $P$ of panels (c), (d).}

At weak rotation, the spectrum traces the intrinsic slope of the background polarization: for Kolmogorov turbulence the 2D spectrum of the LOS-projected polarization scales as $k^{-(2+m_i)} \sim k^{-11/3}$, or $m_i \sim 5/3$ (\citetalias{LazarianPogosyan2012} and \LP), and in the linear regime of Section~\ref{subsec:intrinsic} the directional spectrum inherits this slope, $P_u(k)\propto k^{-11/3}$.  

As $\eta$ increases, the result depends on the imprint of density fluctuations on the RM.  If it is weak, as in the $M_s=1$ simulation, 
and the RM is dominated by the magnetic field, its scaling is similar to the Kolmogorov one and Faraday rotation does not have a pronounced signature.
For high sonic Mach number ($M_s=10$), the density spectrum, and hence the RM-density spectrum, is shallow,
and the projected RM spectrum inherits it, for instance, $m_\Phi \approx 0.42 < 5/3$ in our simulations.
A Faraday-dominated regime appears at high~$k$ reflecting this scaling, with the
transitional scale $k_\times$ shifting to lower~$k$ as predicted by Equation~(\ref{eq:ktimes}).
The limited inertial range of the MHD runs prevents precise two-slope fitting. In Figure~\ref{fig:dirspec_athena} power-law behavior 
is confined between the driving band and the onset of numerical dissipation at $kL_{\rm box}/2\pi\approx80$. Further steepening is numerical rather than physical, and it is the main reason we complement the MHD runs with the higher-resolution synthetic screens of Section~\ref{subsec:synthetic}.

\section{Results of Synthetic Simulations}
\label{sec:results_synthetic}

In this section we analyze the role of fluctuations in the density of thermal electrons within the Faraday screen,
using synthetic simulation cubes described in Section~\ref{subsec:synthetic}.

To isolate the Faraday response and systematically test the two-slope prediction, we consider three configurations:
\begin{enumerate}
\item {Density-only screen.} We fix $B_\parallel=\text{const}$ in the Faraday screen, so that $\Phi\propto\int n_e\,dz$. This separates the Faraday exponent $m_\Phi$ (set by the density spectrum) from the intrinsic exponent $m_i$ (set by the perpendicular magnetic field), and reveals the two-slope regime without ambiguity.
\item {Full screen with only perpendicular mean magnetic field.} We include fluctuations of $B_\parallel$ in the screen, which is more realistic and allows for a direct comparison with the \textsc{AthenaK} results, but mixes the density and $B_\parallel$ spectra in the RM density.
\item {Full screen.} We include fluctuations of $B_\parallel$ in the screen as well as a nonzero parallel component of the mean B-field, which is a general case. The presence of the mean field along the LOS enhances the contribution from the electron density inhomogeneities.
\end{enumerate}

\begin{figure*}[!t]
  \centering
  \begin{minipage}[t]{0.32\textwidth}
    \centering
    \includegraphics[width=\linewidth]{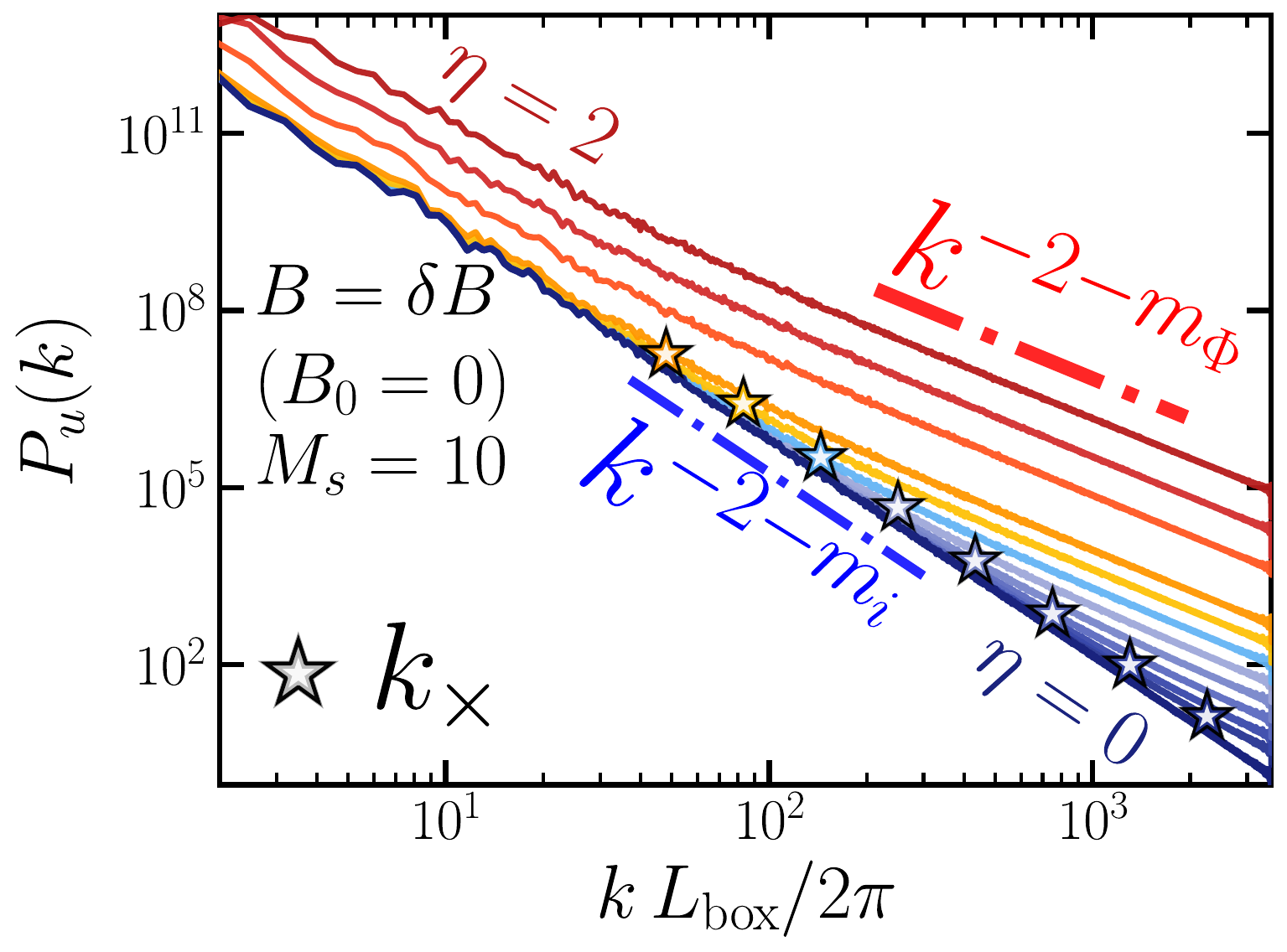}
  \end{minipage}\hfill
  \begin{minipage}[t]{0.32\textwidth}
    \centering
    \includegraphics[width=\linewidth]{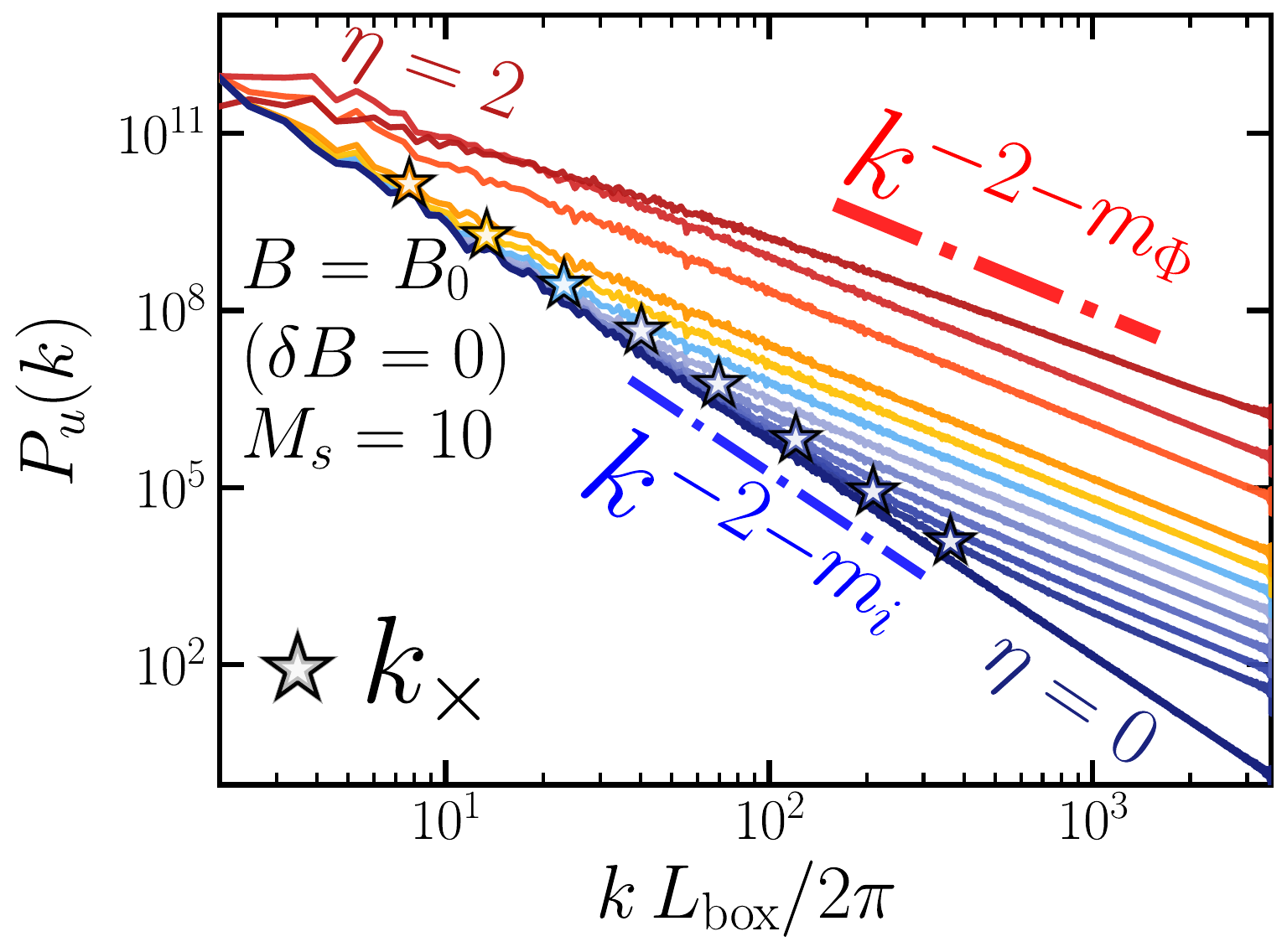}
  \end{minipage}\hfill
\begin{minipage}[t]{0.32\textwidth}
    \centering
    \includegraphics[width=\linewidth]{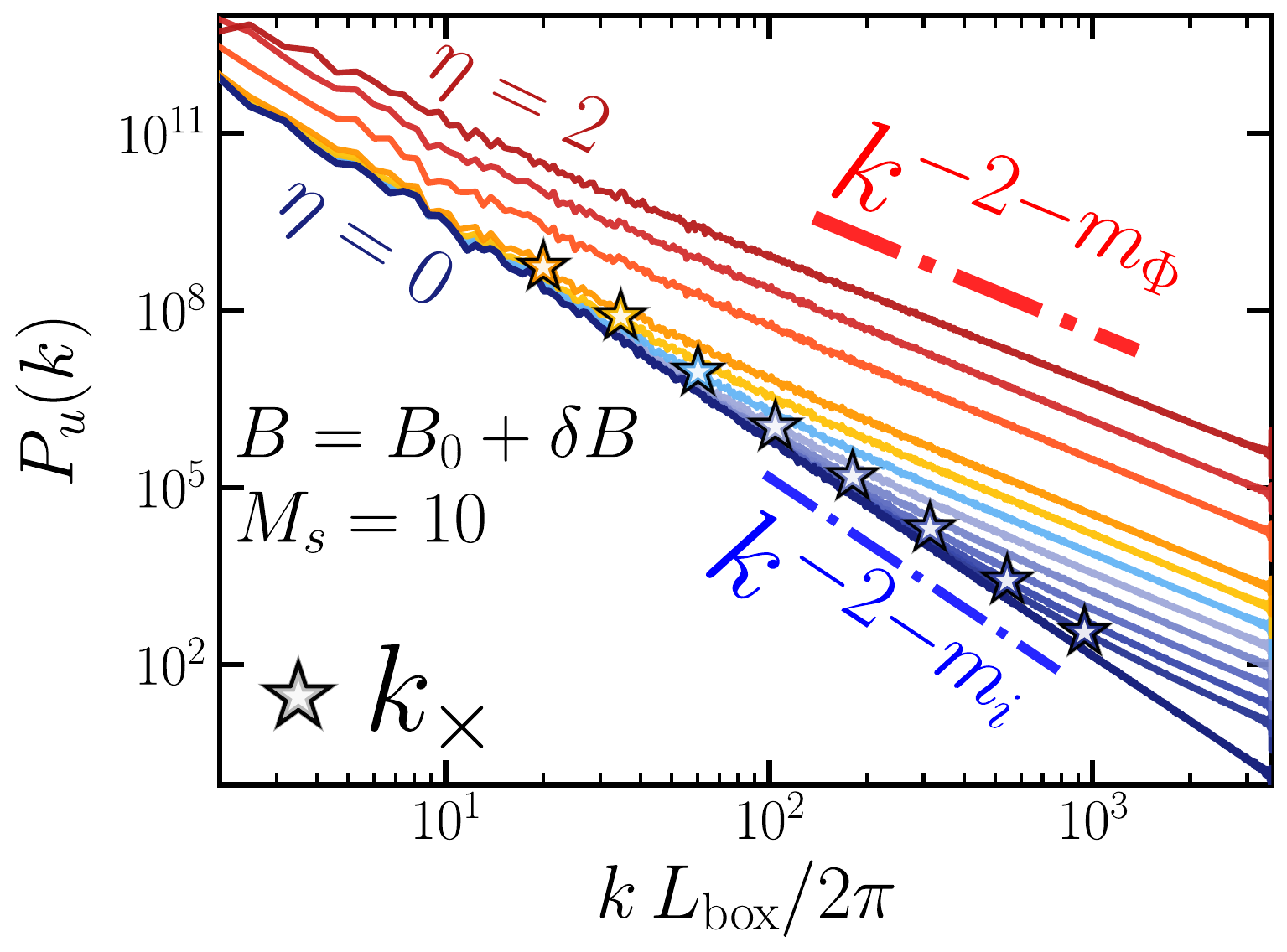}
  \end{minipage}

\centering
  \begin{minipage}[t]{0.32\textwidth}
    \centering
    \includegraphics[width=\linewidth]{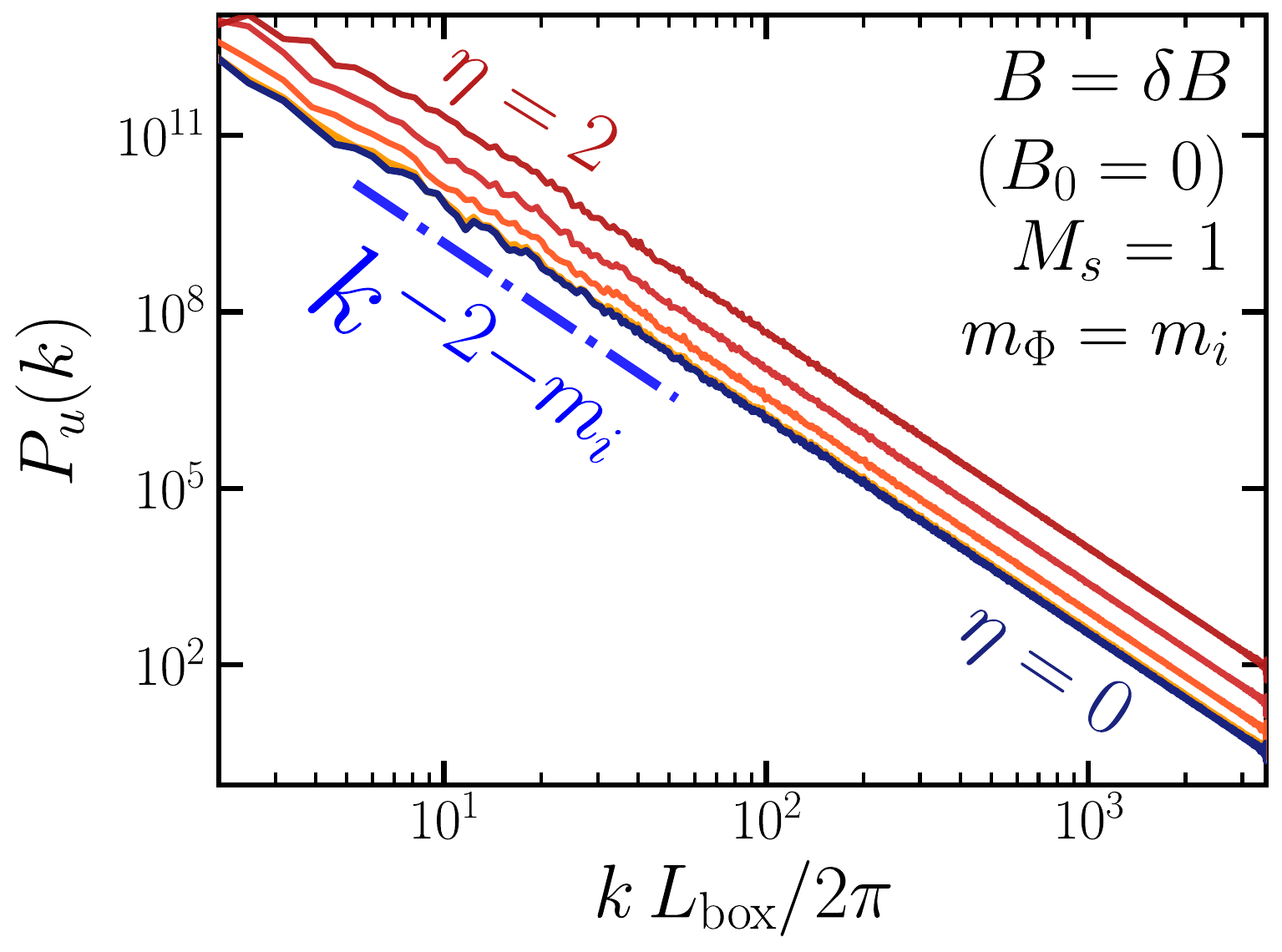}
  \end{minipage}\hfill
  \begin{minipage}[t]{0.32\textwidth}
    \centering
    \includegraphics[width=\linewidth]{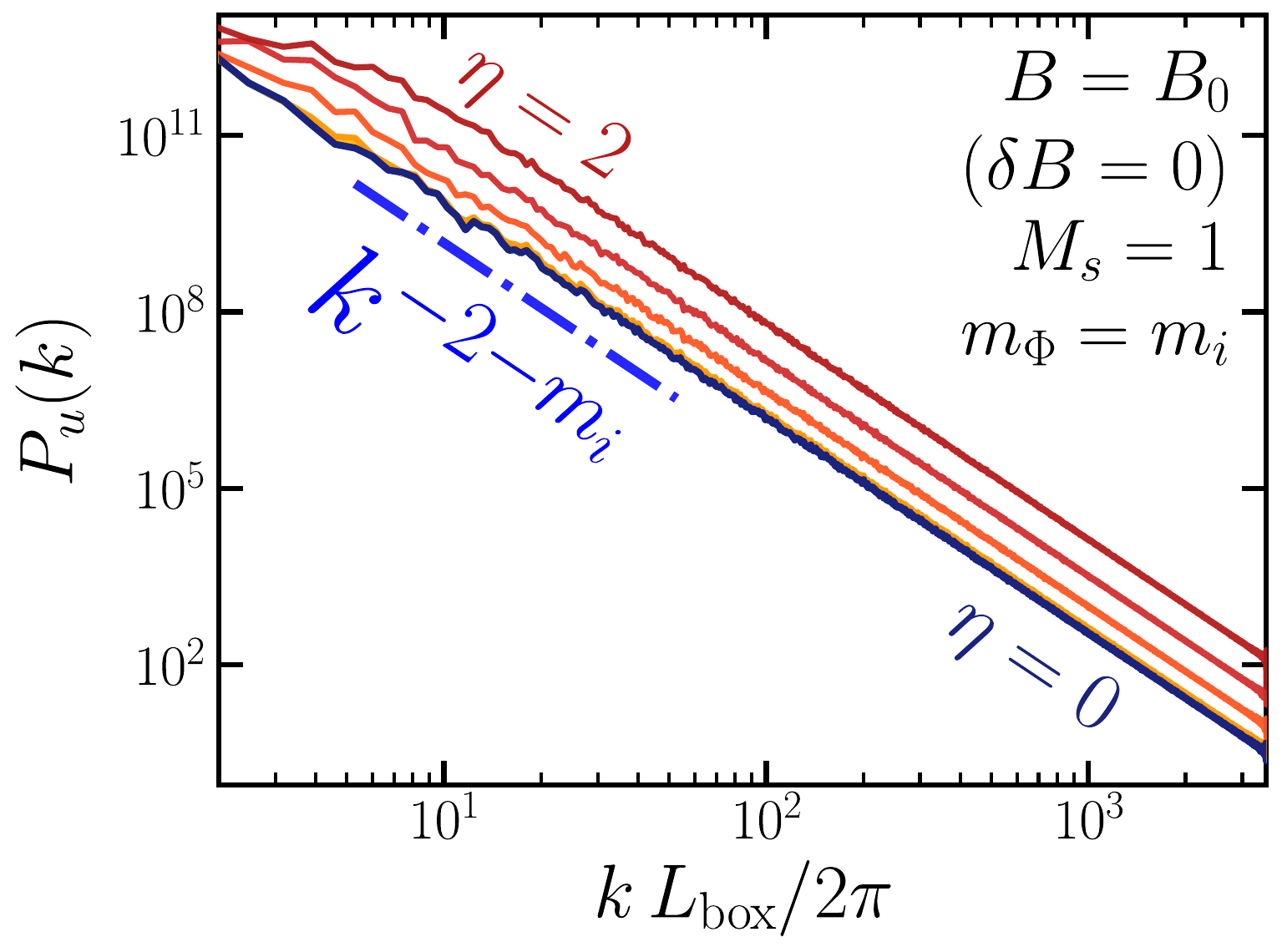}
  \end{minipage}\hfill
\begin{minipage}[t]{0.32\textwidth}
    \centering
    \includegraphics[width=\linewidth]{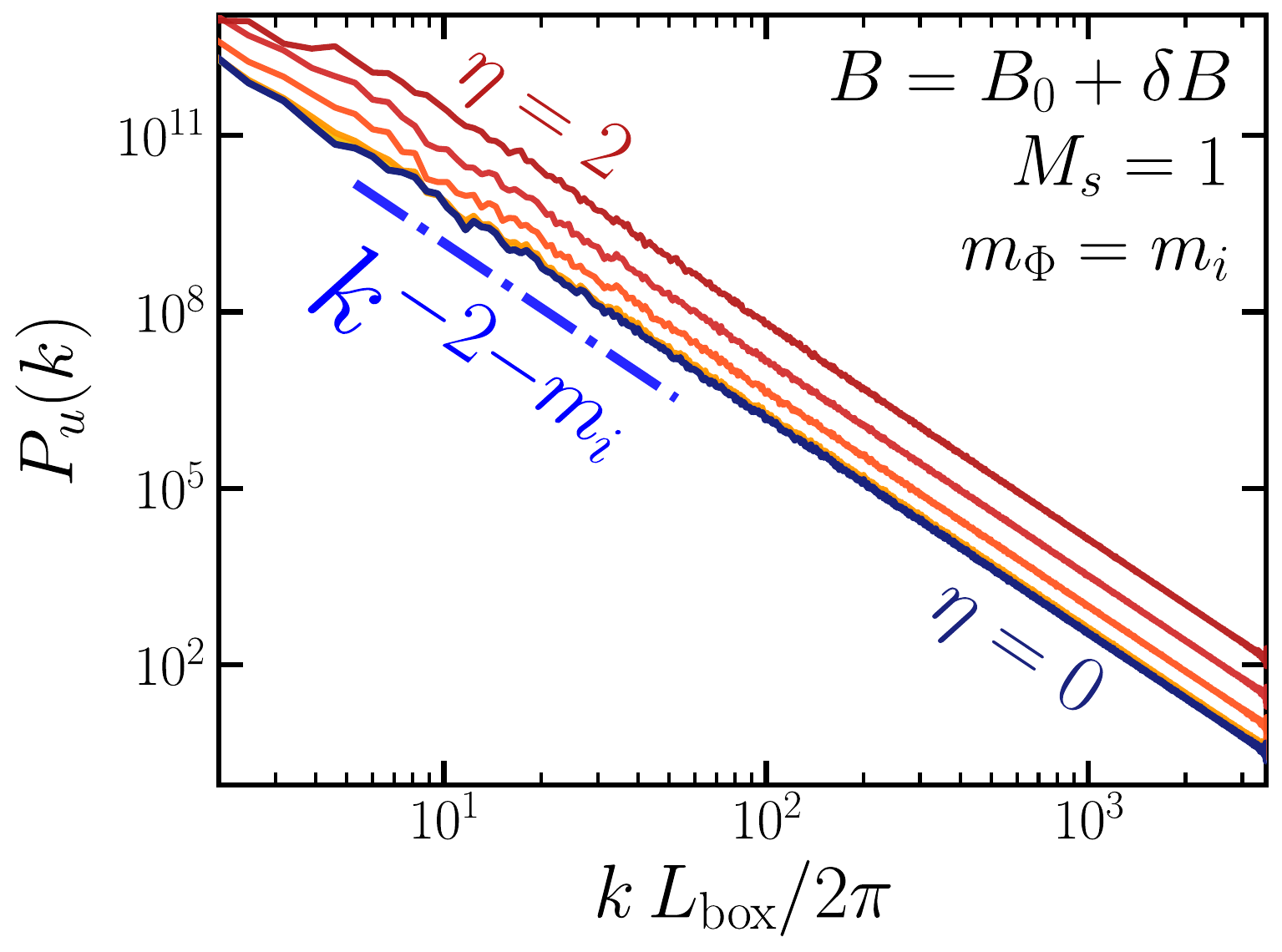}
  \end{minipage}

  \caption{Directional spectrum $P_u(k)$ of the $5040^3$ synthetic separated-screen cubes of Section~\ref{subsec:synthetic}, with $M_s=10$ in the top row and $M_s=1$ in the bottom row. In all panels the RM is $\Phi\propto\int n_e B_\parallel\,dz$ with the same electron density, and the columns differ only in the parallel field of the screen, $B_\parallel=\delta B$ (left, $B_0=0$), $B_\parallel=B_0$ (middle, $\delta B=0$) and $B_\parallel=B_0+\delta B$ with $B_0=2.3\,\sigma_B$ (right). Curves give $\eta=0$ (blue) and 11 values up to $2$ (red), eight of them from $0.01$ to $0.15$ then $0.44$, $0.94$ and $2$. Dashed--dotted lines are the reference slopes of Equation~\eqref{eq:Pu_asymptotics}, offset for clarity and placed where the curves have those slopes, namely, $k^{-2-m_i}$ (blue) with $m_i=5/3$, the prescribed Kolmogorov field of the emitter, on the $\eta=0$ curve, and in the $M_s=10$ row $k^{-2-m_\Phi}$ (red) with $m_\Phi=0.26$, the prescribed $k^{-2.26}$ density spectrum of the screen, on the $\eta=2$ curve. For $M_s=1$ the two indices coincide, so a single line is drawn and the panels are marked $m_\Phi=m_i$. Stars in the $M_s=10$ row mark the crossover $k_\times=C/R_\times$ of Section~\ref{subsec:hankel}, at which the intrinsic and Faraday terms of Equation~\eqref{eq:Pu_asymptotics} are equal. They are computed from the amplitudes $4\sigma_\psi^2 r_i^{-m_i}$ and $r_\phi^{-m_\Phi}$ measured on the input maps, not from the curves, and lie on the bend of every curve with $\eta<2\sigma_\psi$. The three curves above this value carry no star because there the rms Faraday rotation angle $\eta/2$ exceeds the rms intrinsic angle $\sigma_\psi$, so the Faraday term dominates at every scale and no crossover exists. The Faraday branch has the same slope in all three columns and the columns differ in its amplitude, so the stars sit at higher $k$ in the left column, where the screen variance is dominated by the $\bar n_e\,\delta B_\parallel$ term with the $k^{-11/3}$ slope of the source. That term shows up only in the curves with $\eta>2\sigma_\psi$ of the left and right columns, which are steep at large scales and bend to the Faraday slope, whereas in the middle column these curves are shallow at every $k$.}
  \label{fig:dirspec_model}
\end{figure*}

Figure~\ref{fig:dirspec_model} shows the directional spectrum $P_u(k;\lambda)$ measured directly from the synthetic cubes of Section~\ref{subsec:synthetic}.
In the synthetic simulations the inertial-range slope is clearly resolved over more than three decades in~$k$ due to the absence of the short-scale dissipation that limits MHD simulations. This 
allows the different scaling regimes to be observed clearly.

The three columns in Figure~\ref{fig:dirspec_model} isolate different contributions from the magnetic field to the RM $\Phi\propto\int n_e B_\parallel\,dz$. The left column sets the mean field parallel to the LOS to zero and keeps only its fluctuations ($B_\parallel=\delta B$), the middle column holds the parallel field uniform ($B_\parallel=B_0=\mathrm{const}$), so that the RM varies only through the electron density, and the right column includes both ($B_\parallel=B_0+\delta B$). In every column the electron-density fluctuations are retained.  Fluctuations of the perpendicular component of the magnetic field,
$\delta B_\perp$ are the same in all columns as well. The spectral analysis removes the uniform rotation contribution, so $P_u(k>0)$ reflects the fluctuating part of the RM $\delta \Phi \sim \int dz 
\left( \bar n_e \delta B_\parallel + \bar B_\parallel \delta n_e + \delta n_e \delta B_\parallel \right) $ alone. 

For the high level of electron density inhomogeneities, as in the case of $M_s=10$,  the fluctuations of the RM in the Faraday screen dominate the signal at small scales, while the long waves are dominated by intrinsic correlations at the background source with a clear transition at the domain boundary $k_\times$ that depends on $\eta$ (for $\eta<2\sigma_\psi$, see Section~\ref{subsec:hankel}).  In the RM range the details of the magnetic field behavior
in the screen are secondary to the overall picture. Short waves are dominated by density even if there is no mean field in the screen in the parallel direction (left column). In the latter case the effective RM-density spectrum is a convolution of the density and the magnetic field spectra, which lowers the amplitude of the Faraday term and moves the crossover to smaller scales but does not destroy the two-regime structure.
The slopes $m_i$ and $m_\Phi$ remain recoverable, and $R_\times$ continues to follow the scaling of Equation~(\ref{eq:Rtimes}).
On a more careful investigation (compare the curves with $\eta>2\sigma_\psi$ in the central and right panels of the top row), one can see the long-wave contribution of $\delta B_\parallel$ in the RM on top of the background
source contribution. In our model, however, they have the same scaling and are not distinguishable by the slope parameter. Inclusion of 
$\delta B_\parallel$ in the RM also dilutes the amplitude of the short-wave part of the spectrum. 

For $M_s=1$,   density fluctuations are subdominant in the Faraday rotation, and all three panels are dominated by the fluctuations of the magnetic field at the source and within the screen.
With source and screen having the same scaling for the magnetic field, $m_i \approx m_\Phi$, 
we cannot distinguish the contribution of the Faraday rotation.  The only visible effect is a parallel upward shift of the curves that grows with $\eta$ and is similar in the three columns.  
The two-slope regime will be restored if the magnetic fields at the source and the screen have different properties.

\section{Comparison with Alternative Measures in the Separated-screen Regime}
\label{sec:comparison_separated_screen}

\begin{figure*}[!t]
  \centering
  \includegraphics[width=\linewidth]{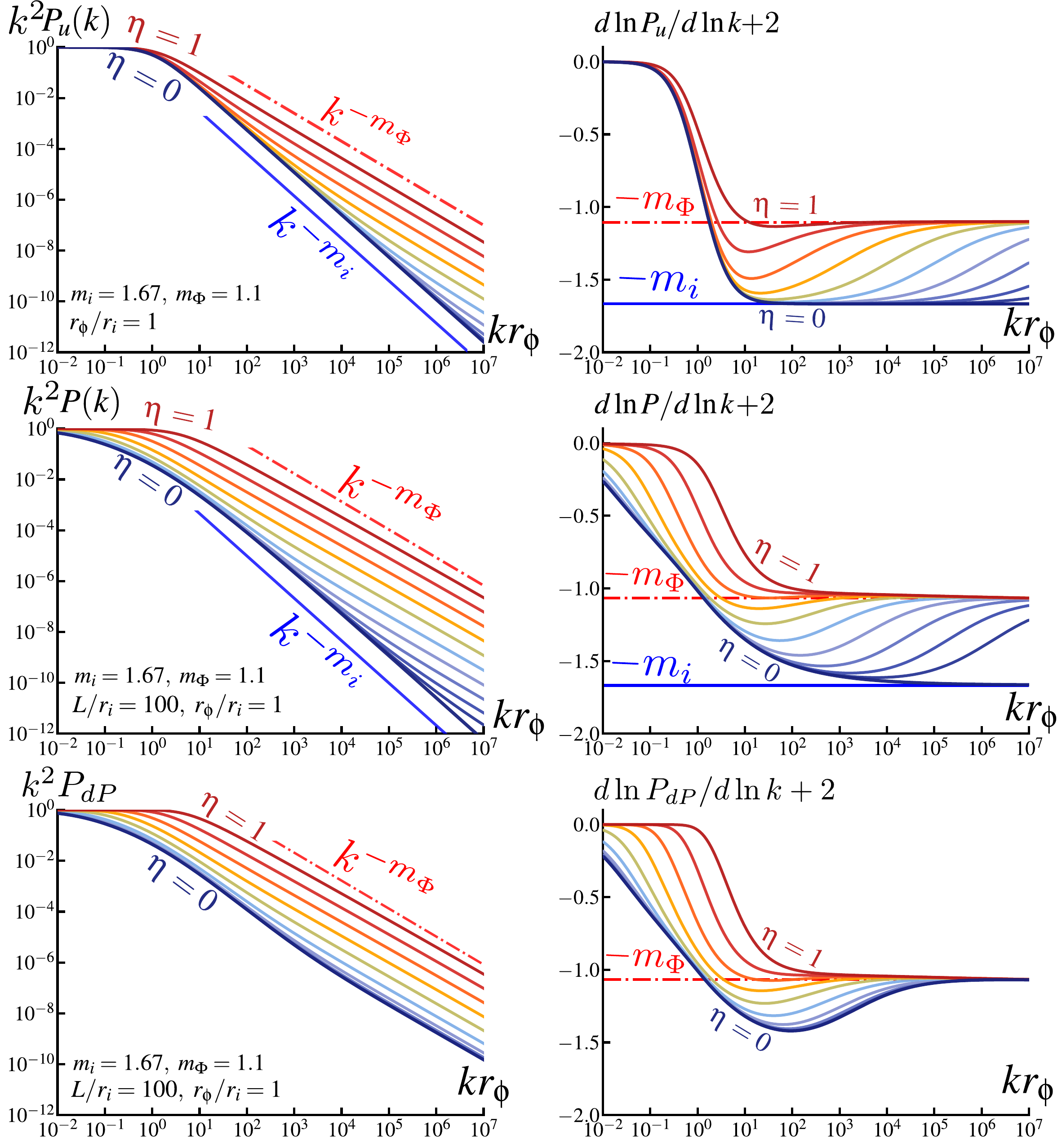}
\caption{Spectral comparison in the separated-screen geometry. The top row shows the
directional spectrum $P_u(k)$ from Equation~\eqref{eq:Pu_def} (left) and its local
logarithmic slope $d\ln P_u/d\ln k + 2$ (right). The middle row shows the \LP{}
polarization spectrum $P(k)$ and its local slope. The bottom row shows the spectrum
$P_{dP}(k)$ of the wavelength-derivative measure $dP/d\lambda^{2}$ and its local slope.
Colored curves correspond to increasing Faraday rotation strength from $\eta=0$ (blue)
to $\eta=1$ (red). The solid blue and dashed--dotted red guide lines mark the inertial-range scalings $k^{-m_i}$ and
$k^{-m_\Phi}$ (here $m_i=5/3$, $m_\Phi=1.1$, $r_\phi/r_i=1$). All three measures involve
the same two inertial-range slopes, the intrinsic $-m_i$ and the Faraday $-m_\Phi$, but
they approach them differently. $P_u$ and $P$ run from the intrinsic slope $-m_i$ at
$\eta=0$ to the Faraday slope $-m_\Phi$ at $\eta=1$ ($P_u$ reaching $-m_\Phi$ already at
moderate $\eta<1$ across more than two decades in $k$, while $P$ retains a mixed slope
over much of the inertial range at the same $\eta$). The derivative $dP/d\lambda^2$
instead keeps a Faraday-dominated branch ($-m_\Phi$) at small scales (large $k$) even at
$\eta=0$, with a steeper, intrinsic-influenced branch (approaching but not reaching
$-m_i$) at larger scales, at the cost of requiring multifrequency data. Here the
small-$\eta$ limit corresponds to the short-wavelength limit at fixed RM dispersion
$\sigma_\Phi$ (the screen retained), not to $\sigma_\Phi=0$, for which the derivative
signal vanishes, $\lim_{\sigma_\Phi\to0}\langle|\partial_{\lambda^2}P|^2\rangle=0$.}
  \label{fig:spec_compare}
\end{figure*}

To place our angle-based directional spectrum in context, we compare it with the two spatial diagnostics that \LP{} derive for the separated-screen geometry (see Appendix~C of \LP{}): the polarization structure function $D_P(R)$ and the $dP/d\lambda^2$ correlation. The spectral comparison is shown in Figure~\ref{fig:spec_compare}, and the real-space (structure function) counterpart is given in Figure~\ref{fig:sf_compare} in the Appendix.

A first look at Figure~\ref{fig:spec_compare} shows that all three measures share the same inertial-range endpoints but differ in how rapidly the local slope migrates between them as $\eta$ is increased. For $P_u$ (top row) the transition is essentially complete by $\eta \sim 1$, with the slope reaching $-m_\Phi$ across more than two decades in $k$. For the \LP{} polarization spectrum $P$ (middle row) the same $\eta=1$ curve still carries an intermediate slope across much of the inertial range. The $\lambda^4 D_\Phi$ suppression in the amplitude-weighted correlator delays the transition. The derivative $dP/d\lambda^{2}$ (bottom row) tracks $-m_\Phi$ from small $\eta$ but does so over a narrower range of $k$, and constructing it requires a clean multifrequency sampling. The three diagnostics are therefore complementary, trading off single-band amplitude robustness ($P_u$), high-frequency emitting region constraints ($P$, $D_P$), and multiband Faraday amplification ($dP/d\lambda^{2}$).
\subsection{Polarization Structure Function}
\label{subsec:comp_P}

In the separated-screen geometry, the polarization correlation factorizes (see Equation~(\ref{eq:xiu_factorized})) as
\begin{equation}
\langle P(\bm X)\,P^\ast(\bm X+\bm R)\rangle
=\xi_{P_i}(\bm R)\;\exp[-2\lambda^4 D_\Phi(R)]\,,
\end{equation}
where $\xi_{P_i}$ is the {full} (amplitude-weighted) intrinsic polarization correlation.  As a result, the correlation statistics are heavily weighted toward the regions with high polarization, losing the directional information from low-polarization lines of sight.
By contrast, our directional spectrum $P_u$ is based on the amplitude-normalized field $u=P/|P|$ and responds more directly to Faraday rotation 
by focusing on the direction of the polarization.
This makes $P_u$ sensitive to the Faraday screen over a wider range of wavelengths.

\subsection{Wavelength Derivative $dP/d\lambda^2$}
\label{subsec:comp_dP}

The derivative of the polarization field with respect to the (square of the) wavelength is another statistic, suggested by \LP. 
From
\begin{equation}
\frac{dP}{d\lambda^2}=2i\,\Phi(\bm X)\,P(\bm X,\lambda^2)\,,
\label{eq:dP_dlambda2}
\end{equation}
we see that taking the derivative explicitly weights the signal by the foreground RM, reflecting the RM factor even at very short wavelengths $\lambda$.
\LP{} showed that in the weak-decorrelation regime the structure function of $dP/d\lambda^2$ is sensitive to Faraday rotation without the $\lambda^4$ suppression that limits $D_P$. This makes $dP/d\lambda^2$ an excellent Faraday amplifier.

This distinctive feature of the wavelength-derivative measure is manifest in the limit of weak differential rotation $\lambda^4 D_\Phi(R) \ll1$. 
Following \LP{},
in this regime the derivative correlation
$\xi_{dP}(R) \equiv \left\langle \partial_{\lambda^2}P(\bm X)\,\partial_{\lambda^2}P^*(\bm X{+}\bm R)\right\rangle $
reduces to the product of the intrinsic polarization
correlation $\xi_{P_i}$ and the RM correlation $\xi_\Phi(R)$,
\begin{equation}
\xi_{dP}(R) \approx 4\,\xi_{P_i}(\bm R)\,\xi_\Phi(R),
\label{eq:dPdl2_weak}
\end{equation}
whereas the corresponding
structure function $D_{dP}(R)\equiv\langle|\partial_{\lambda^2}P(\bm X)-\partial_{\lambda^2}P(\bm X{+}\bm R)|^2\rangle$ becomes a sum of the source and Faraday terms,
\begin{equation}
D_{dP}(R)
\approx 4\sigma_\Phi^2\,D_{P_i}(R)+4\,\xi_{P_i}(\bm R)\,D_\Phi(R),
\label{eq:DdP_weak}
\end{equation}
where $D_{P_i}(R)=\langle|P_i(\bm X)-P_i(\bm X+\bm R)|^2\rangle$ and $\xi_\Phi(R)=\sigma_\Phi^2-D_\Phi(R)/2$.
The right-hand sides of
Equations~\eqref{eq:dPdl2_weak}--\eqref{eq:DdP_weak} carry no explicit $\lambda$, so the RM
structure function $D_\Phi(R)$ enters $D_{dP}$ without the $\lambda^4$ suppression that limits $D_P$ itself, and survives even in the short-wavelength limit. Consequently
$dP/d\lambda^2$ retains a two-slope structure even at small $\eta$, and which contribution
dominates on a given scale is set by the relative exponents $m_i$, $m_\Phi$ and
correlation lengths $r_i$, $r_\phi$, not by $\lambda$. The Faraday-dominated branch
($k^{-m_\Phi}$ in the spectrum, $R^{m_\Phi}$ in real space) appears at the smallest
scales, where the short RM correlation length $r_\phi$ dominates, while a steeper,
intrinsic-influenced branch (tending to $m_i$) takes over at larger scales. The
limited dynamic range available before the structure function saturates means the
intrinsic branch does not fully reach $m_i$ in
Figures~\ref{fig:spec_compare} and \ref{fig:sf_compare}, but the small-scale Faraday branch
is resolved at every $\eta$.

We emphasize that the weak-rotation limit considered here is reached by decreasing the
wavelength while keeping the Faraday screen
in place with finite RM dispersion $\sigma_\Phi$. It is not the limit of a vanishing screen, $\sigma_\Phi=0$. The two are
physically distinct. Removing the screen sends $D_\Phi$, and hence the entire derivative signal,
to zero (its amplitude collapses and $r_\phi$ is no longer defined), whereas shortening
the wavelength leaves the RM statistics intact and merely switches off the $\lambda^4$
damping.

However, the derivative measure comes at an observational cost. Estimating $dP/d\lambda^2$ requires accurate multifrequency measurements and can amplify noise (\citetalias{LazarianPogosyan2016}; \citealt{Zhang2016}).
Our angle-based statistic can be constructed from a single $(Q,U)$ map and evaluated via a single FFT (Equation~(\ref{eq:Pu_def})).

For the separated-screen regime with moderate Faraday rotation, the directional spectrum shows greater sensitivity to the Faraday screen slope at fixed wavelength, because it directly traces angle differences rather than amplitude-weighted correlations.
The $dP/d\lambda^2$ measure has comparable or better sensitivity in the strong-rotation regime, but requires multifrequency data.

\section{Discussion and Summary}
\label{sec:conclusions}
\subsection{Directional Spectrum}

Our work extends the theoretical study in \LP{} where obtaining the statistics of turbulence from polarized synchrotron radiation was considered. In our setting the synchrotron radiation arises in a turbulent volume and experiences Faraday rotation in a separate volume \citep[as in the numerical tests of][]{Zhang2018}, and the new element is the amplitude-normalized director statistic. 

The principal advance of this work is the demonstration that the inertial-range statistics of a foreground Faraday screen can be recovered from the polarization directions alone, in a single radio-frequency polarization map.  We employ the directional spectrum $P_u$ given by Equation~(\ref{eq:Pu_def}) which is constructed from the Fourier transform of the Stokes field normalized by amplitude. The spectrum inherits the $\pi$-periodicity of the polarization angle, avoids branch-cut ambiguities, and can be computed with a single fast Fourier transform.  Both our \textsc{AthenaK} MHD simulations and the controlled synthetic-screen experiments confirm that this estimator can reliably trace the statistics of the underlying magnetic turbulence.  For weak Faraday rotation, $P_u$ reproduces the intrinsic synchrotron slope. As the rotation strength $\eta$ increases, a shallower Faraday-dominated regime emerges at high wavenumber, and the transition scale $k_\times$ shifts to lower $k$ in agreement with the analytical prediction of Equation~(\ref{eq:ktimes}).

\subsection{Comparison with Earlier Diagnostics}

We compared $P_u$ with alternative measures explored in \LP.  In particular, \LP{} studied the Faraday rotation statistics using $dP/d\lambda^{2}$. Taking the derivative of the polarization acts as a powerful amplifier of the Faraday contribution. The corresponding structure function is sensitive to Faraday rotation even in the weak-decorrelation regime. In contrast, \LP{} showed that the $\lambda^{4}$ suppression limits the utility of the standard polarization structure function $D_P$.  The limitations of $dP/d\lambda^{2}$ arise from the requirement of low-noise multifrequency observational data that, after differentiation, are not dominated by noise (\citetalias{LazarianPogosyan2016}; \citealt{Zhang2016}).    

By contrast, the directional spectrum $P_u$ employs single-frequency data and does not require differentiation. Our comparison of both the structure functions and spectra of the newly introduced measure with the previous ones (Figure~\ref{fig:spec_compare}; see also Figure~\ref{fig:sf_compare} in the Appendix) shows that, at moderate Faraday rotation, defined as $\eta = 2\sigma_\Phi\lambda^{2} < 1$ in the rotation-strength parameter of Equation~\eqref{eq:faraday_rotation_strength}, equivalently $\lambda^{4} D_\Phi(R) < 1$ across the inertial range, the directional spectrum exhibits greater sensitivity to the Faraday statistics compared to the $dP/d\lambda^{2}$ measure. The difference arises due to $P_u$ directly tracing angle differences rather than amplitude-weighted correlations.  In physical terms, amplitude normalization amplifies the Faraday rotation effect.  The $dP/d\lambda^{2}$ measure becomes advantageous only for $\eta \geqslant 1$, where the Faraday factor dominates the full polarization signal.

The directional spectrum can also be applied across multiple frequency bands when these are available \citep[see the analysis of multifrequency survey cubes by][]{Xiao2026}. The crossover scale $k_\times(\lambda)$ shifts with wavelength according to Equation~\eqref{eq:ktimes}, and tracking this shift between bands provides an independent constraint on $m_\Phi - m_i$ that does not rely on any single-band slope fit. Multifrequency data therefore upgrade $P_u$ from a single-band slope estimator into a joint estimator of the intrinsic and Faraday exponents, without the differentiation step that limits $dP/d\lambda^{2}$ at realistic noise levels. 

We view the directional spectrum $P_u$ as complementary to existing measures.  Indeed, the polarization structure function $D_P$ constrains the emitting-region statistics at high frequency, where Faraday effects are negligible.  The derivative $dP/d\lambda^{2}$ is a way to amplify the effects of turbulent Faraday rotation. The directional spectrum provides an amplitude-independent, single-band probe that is straightforward to implement and that responds to the Faraday screen across a broad range of rotation strengths.

\subsection{Observational Criteria and Practical Considerations}

A key practical outcome of this work is a simple, observable criterion that guides frequency requirements.  The behavior of the bounded angle structure function $D_u(R;\lambda)=2[1-e^{-2\lambda^4 D_\Phi(R)}]$ is controlled by the product $\lambda^4 D_\Phi(R)$. When this quantity is small across the inertial range, $D_u$ linearizes to $\lambda^4 D_\Phi$ and the turbulence slope is directly readable from a single map. When it approaches unity at small scales, the exponential saturates and the slope is difficult to obtain.  In the transitional regime, two nearby frequency bands suffice to bracket the saturation and restore the correct spectral index.  Only when the entire inertial range is saturated is full multifrequency synthesis required.

An observer can therefore adopt the following strategy. Compute the directional spectrum from an existing single-band $(Q,U)$ map, check whether the measured slope steepens toward the saturation, and, if it does, add one adjacent band to interpolate through the transition.  This approach significantly lowers the data requirements compared with derivative-based methods and is well suited to the separated-screen geometry common in Galactic and extragalactic radio polarimetry.

\subsection{Limitations and Future Work}

Our separated-screen geometry neglects the internal depolarization that arises when emission and rotation take place in the same volume.  Extending the directional-spectrum formalism to the mixed case treated by \LP{} \citep[and explored with two-wavelength polarization differences by][]{Hu2025} is a natural next step, particularly for low-frequency observations where internal Faraday rotation is strongest.  We have also assumed Gaussian statistics for the RM increments (Equation~(\ref{eq:xiu_factorized})). This is well justified when the magnetic field dominates the RM fluctuations \citepalias{LazarianPogosyan2016}, whereas the intermittent, shock-dominated densities of highly supersonic screens (our $M_s=10$ case) may produce non-Gaussian RM increments \citep{Seta2023, Wang2024} whose effect on the exponential damping factor warrants further study.

A useful intermediate case, bridging the fully separated geometry studied here and the fully mixed geometry of \LP, is that of two Faraday screens with some emission behind both and some between them. As noted by \LP{}, such intermediate cases can be represented as a superposition of the two limiting cases. Qualitatively, the results obtained here carry over as follows. Any RM common to all the emission along a LOS, i.e., coming from the truly foreground Faraday screen closest to the observer, multiplies the entire signal and factorizes out of the director correlation exactly as the single screen analyzed above, so its statistics are recovered without modification. 
The total emission from the two emitting layers with the second rotating volume in between serves as the background source. Its total spectral properties
can be determined. In fact, for our formalism it does not matter what the origin of the synchrotron spectrum of the background region is,
but the mix of individual terms from separate emitting regions is additive rather than factorized and needs to be studied 
similarly to the mixed-volume situation of \LP{}. 
A quantitative treatment of the fully mixed case, and of multiple interleaved layers, requires the coupled emissivity--rotation formalism of \LP{} and is left to future work.  

Despite these limitations, the present results establish the directional spectrum as a practical, single-frequency diagnostic of magnetized turbulence.  Its minimal data requirements and robustness to amplitude calibration make it well suited for application to current facilities and to forthcoming surveys with the Square Kilometre Array and its precursors \citep[see][]{Beck2015, Haverkorn2015, Akahori2018, Heald2020}.

\subsection{Summary}

Our main results can be summarized as follows.

\begin{enumerate}
\item We explored the utility of a new measure, the directional spectrum, constructed from the amplitude-normalized Stokes field, and demonstrated how it can recover the inertial-range scaling of a foreground Faraday screen using a polarization map measured at a single radio frequency. This opens a new way to study MHD turbulence with the Faraday effect.

\item Using both \textsc{AthenaK} MHD simulations and synthetic data to represent Faraday screens, we tested the analytical criteria for the directional spectrum transition from the intrinsic synchrotron slope at weak Faraday rotation to a shallower Faraday-dominated regime at moderate-to-strong rotation.

\item Compared to the wavelength-derivative diagnostic introduced in \LP, the directional spectrum exhibits greater sensitivity to the Faraday screen for moderate Faraday rotation, as amplitude normalization suppresses the effects of intrinsic brightness variations and directly exposes the Faraday-angle imprint. The two measures are complementary. The derivative requires high signal-to-noise multifrequency data (\citetalias{LazarianPogosyan2016}; \citealt{Zhang2016}), while the directional spectrum can operate with single-frequency data at a higher noise level. 

\item The directional spectrum is also useful with multifrequency data. We provided an observable criterion, the Faraday rotation strength, based on the product of the rotation dispersion and the squared wavelength, that indicates when a single band suffices for slope recovery, when two nearby bands are needed in the transitional regime, and when full multifrequency synthesis is required.

\item Our results open a practical pathway to extract three-dimensional magnetic turbulence statistics from existing and forthcoming single-band radio polarimetry, thereby lowering data requirements relative to multifrequency techniques and complementing them in a synergistic manner.
\end{enumerate}

\section*{Acknowledgments}
A.L. acknowledges the support of NSF grants AST 2307840 and NASA SEP 80NSSC25K0067. We thank Yue Hu for consultations related to the numerical simulations and helpful suggestions. A.M. acknowledges research and conference-travel support from Berea College.

\section*{\texorpdfstring{Appendix\\ Structure Function for Polarization Directions}{Appendix: Structure Function for Polarization Directions}}
\setcounter{figure}{0}
\renewcommand{\thefigure}{A\arabic{figure}}
\renewcommand{\theHfigure}{A\arabic{figure}}

\begin{figure*}[!t]
  \centering
  \includegraphics[width=\linewidth]{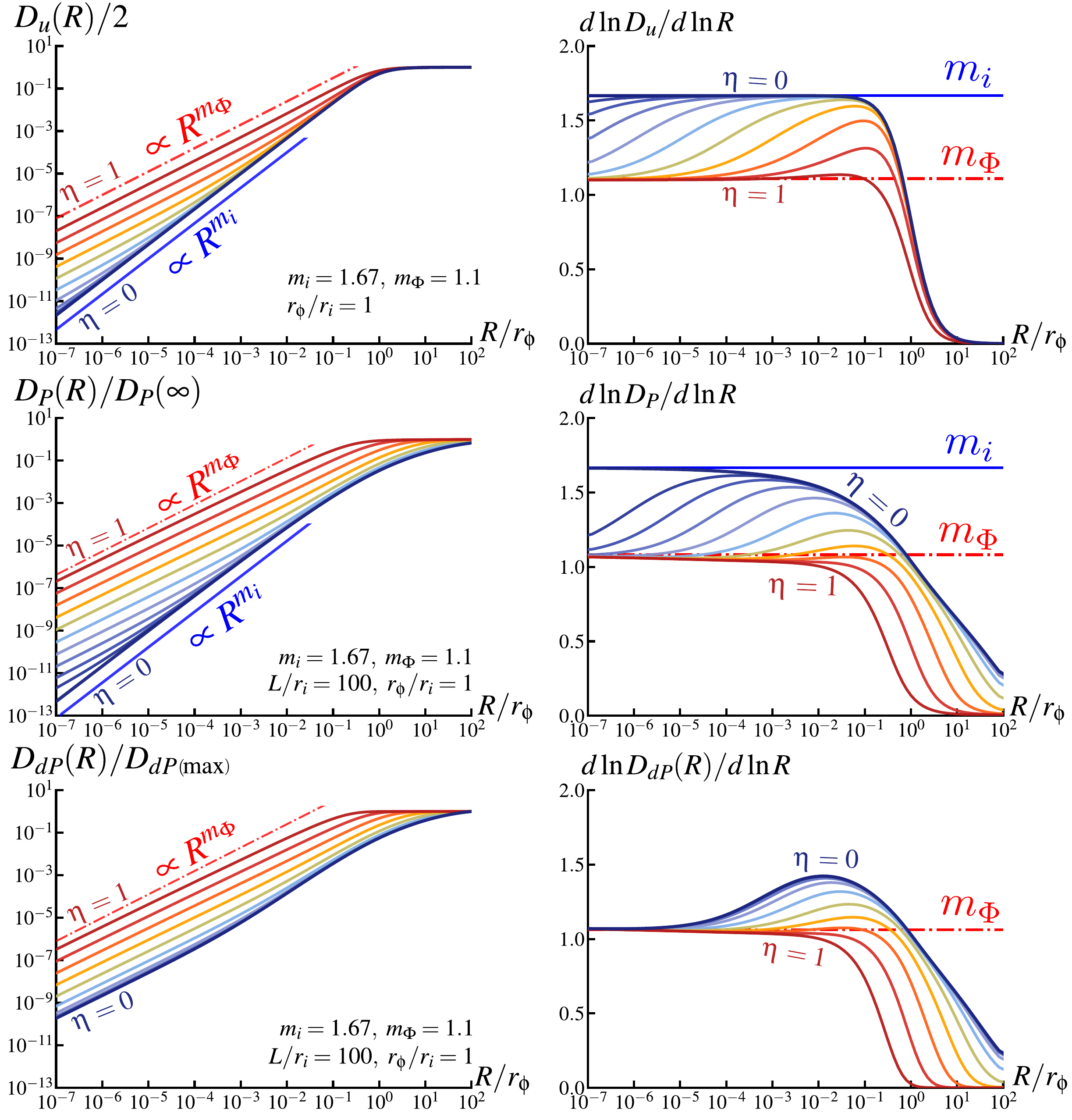}
  \caption{Comparison of the structure function of the directional measure with the measures introduced in \LP{} in the separated-screen model for increasing Faraday rotation strength $\eta$ (colored curves, from blue at the smallest to red at the largest $\eta$, with $\eta=0$ and 10 logarithmically spaced values from $3\times10^{-3}$ to $1$). The \textit{left column} shows the structure functions, and the \textit{right column} shows their local logarithmic slopes $d\ln(\cdot)/d\ln R$. The top row shows the directional structure function $D_u(R)/2$ from the weak-rotation ansatz (Equation~(\ref{eq:du})). The middle row shows the normalized polarization structure function $D_P(R)/D_P(\infty)$ computed with the \LP{} kernel formalism. The bottom row shows the \LP{} derivative-based measure, shown as the normalized structure function $D_{dP}(R)/D_{dP,\max}$ (constructed from $dP/d\lambda^2$). In all rows, the blue solid and red dotted--dashed guide lines mark the expected inertial-range scalings $\propto R^{m_i}$ and $\propto R^{m_\Phi}$, respectively (parameters as in Figure~\ref{fig:spec_compare}, with $R$ plotted in units of $r_\phi$). As $\eta$ increases, the Faraday contribution expands and the slope transitions from the intrinsic asymptotics ($m_i$) to the Faraday screen asymptotics ($m_\Phi$), making the analysis of the structure function $D_u$ complementary to the study of the directional spectrum $P_u$. In the bottom row, $D_{dP}$ keeps a Faraday slope ($m_\Phi$) branch at small $R$ for
all $\eta$, including small $\eta$.}
  \label{fig:sf_compare}
\end{figure*}

Based on the analytical prediction of the structure function $D_u/2$ in Equation~\eqref{eq:du}, we extend the spectral comparison of Figure~\ref{fig:spec_compare} to real space. Figure~\ref{fig:sf_compare} summarizes the real-space comparison of the three diagnostics discussed in the main text: the director-field structure function $D_u(R;\lambda)/2$, the \LP{} polarization structure function $D_P(R)/D_P(\infty)$, and the derivative-based measure $D_{dP}(R)/D_{dP,\max}$. For each statistic, the left panel shows the structure function and the right panel shows the local logarithmic slope $d\ln(\cdot)/d\ln R$ for increasing Faraday rotation strength $\eta$ (blue to red). The reference lines with slopes $m_i$ and $m_\Phi$ highlight the transition from the intrinsic-dominated regime to the Faraday-dominated regime as $\eta$ increases. The real-space picture confirms the spectral comparison of Figure~\ref{fig:spec_compare}. The local slope of $D_u$ reaches the Faraday screen asymptote $m_\Phi$ already at moderate $\eta$, while $D_P$ requires stronger rotation and $D_{dP}$ tracks $m_\Phi$ from small $\eta$ at the cost of needing multifrequency input.

\clearpage
\onecolumngrid
\begin{center}ORCID iDs\end{center}
\noindent A. Melnichenka\,\includegraphics[width=9pt]{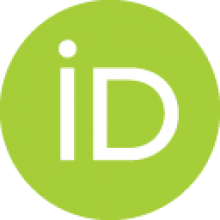} \url{https://orcid.org/0009-0004-1969-779X}\\
A. Lazarian\,\includegraphics[width=9pt]{Orcid-ID.png} \url{https://orcid.org/0000-0002-7336-6674}\\
D. Pogosyan\,\includegraphics[width=9pt]{Orcid-ID.png} \url{https://orcid.org/0000-0002-7998-6823}

\bibliographystyle{aasjournalv7}

\bibliography{refs}
\label{lastpage}

\end{document}